\documentclass[aps,pre,reprint,superscriptaddress,longbibliography,showpacs,floatfix]{revtex4-1} 
\usepackage{graphicx}

\usepackage{times,bm,url,amsmath,upgreek}
\usepackage{amssymb}
\usepackage[colorlinks=true,breaklinks=true,allcolors=blue]{hyperref}
\usepackage[utf8]{inputenc}

\graphicspath{ {./figures/} }
\newcommand{\orcidicon}[1]{\href{https://orcid.org/#1}{\includegraphics[height=\fontcharht\font`\B]{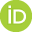}}} 

\usepackage{comment}

\usepackage[x11names]{xcolor}

\newcommand{\mat}[1]{\begin{bmatrix}#1	\end{bmatrix}}

\def\Oo{\ensuremath{{\cal O}}} 
\def\Qq{\ensuremath{{\cal Q}}} 
\def\Hh{\ensuremath{{\cal H}}} 
\def\Oo{\ensuremath{{\cal O}}} 
\def\Ff{\ensuremath{{\cal F}}} %
\def\Rr{\ensuremath{{\cal R}}} 
\def\Jj{\ensuremath{{\cal J}}} 
\def\Ll{\ensuremath{{\cal L}}}
\def\Ww{\ensuremath{{\cal W}}} 
  
\def\Uu{\ensuremath{{\cal U}}}  
\def\Tt{\ensuremath{{\cal T}}}  
  
\def\Ss{\ensuremath{{\cal S}}} 
\def\Hh{\ensuremath{{\cal H}}} 
\def\Nn{\ensuremath{{N}}}
\def\p{\ensuremath{\partial}}

\def\i{\ensuremath{\imath}}

\usepackage[scr=boondox]{mathalfa}

\usepackage{xparse}
\DeclareDocumentCommand{\nint}{ O{} O{} m }{\ensuremath{ \int_{\mbox{\scriptsize $#1$}}^{\mbox{\scriptsize$#2$}}\!\!\! \mbox{\small $\,\mathrm{d}#3$\! }}}

\usepackage{bbold} 
\def\1{\ensuremath{{\mathbb{1}}}}

\begin{document}
	\title{Matching partition functions of deformed JT gravity and  cSYK model}
	
	\author{Jan C.\ Louw \!\orcidicon{0000-0002-5111-840X}}
	\affiliation{Institute for Theoretical Physics, Georg-August-Universit{\"a}t G{\"o}ttingen,  Friedrich-Hund-Platz 1, 37077~G{\"o}ttingen, Germany}
	\author{Sizheng Cao}
	\affiliation{Department of Physics, College of Sciences, Shanghai University, 99 Shangda Road,
200444 Shanghai, China}
	\author{Xian-Hui Ge}
	\affiliation{Department of Physics, College of Sciences, Shanghai University, 99 Shangda Road,
200444 Shanghai, China}
 \affiliation{Shanghai Key Laboratory of High Temperature Superconductors, Shanghai University, 99 Shangda Road, Shanghai 200444, China}
	\date{\today}
	\begin{abstract}
		Motivated by recent analogies between the large-$q$ cSYK model and charged black holes, we aim to find a concrete gravitation theory with a matching partition function. Our main focus is to match the thermodynamics of the $(0+1)$-dimensional cSYK model, with that of a $(1+1)$-dimensional gravitational model. We focus on a model of deformed JT gravity, characterized by some unknown dilaton potential function and unknown dilaton-to-Maxwell field coupling. By finding the general solutions, we are able to find the Lagrangian which produces the same partition function and equation of state as that of the considered SYK model. We go beyond showing that the thermodynamics overlaps by also showing that the Lyapunov exponents, characterizing the degree of chaos, overlap close to the second-order phase transition. In the low-temperature rescaled regime, there remain open questions about the Lyapunov exponents, given that our analysis ignores the black hole back action which can be large in this regime. 
	\end{abstract}

	\maketitle
	\section{Introduction and outline}
   The Sachdev-Ye-Kitaev (SYK) model is a simple quantum model that proposes a gravity-condensed matter correspondence. One of its key findings is the emergence of conformal symmetry with nearly AdS$_2$ geometry in its configuration space of reparametrization modes \cite{Sarosi2018},  which is also observed in black holes. Both systems are also maximally chaotic \cite{Kitaev2015,Maldacena2016}. Significant progress has been made in understanding this duality, including the discovery that fluctuations away from conformality are described by a Schwarzian action \cite{Trunin2021Jun}, which is also the boundary theory of Jackiw-Teitelboim (JT) gravity.  There is a wealth of literature on the connections between the SYK models and JT gravity \cite{Anninos2021,Anninos2022,Trunin2021Jun,Maldacena2016,Cao2021Mar,Chowdhury2022,Kitaev2018,Almheiri2015,Maldacena2016b,Jensen2016,Mandal2017,Lala2019,Lala2020}. The chaotic-integrable transition in the SYK model can be achieved by introducing a generalized SYK model with an additional one-body infinite-range random interaction \cite{Antonio2018Jun}. This transition is interpreted as the Hawking-Page (HP) phase transition in the bulk gravity \cite{Antonio2018Jun}. 
   
   Attempts have been made to extend such holographic analogies to charged black holes by considering complex SYK (cSYK) models \cite{Louw2023}. The cSYK model exhibits a second-order phase transition in the maximally chaotic regime, which is believed to be associated with a universal class of phase transitions in spherical Reissner-Nordstr\"om (RN)-anti-de Sitter (AdS) black holes \cite{Kubiznak2012Jul,Kubiznak2017Feb}. On a thermodynamic level, analogies have been drawn between RN black holes and van der Waals liquid-gas phase transition, and recently also to the phase transition found in the cSYK models \cite{Kubiznak2012Jul,Louw2023}.  Similar phase transitions can be found in (1 + 1)-dimensional deformed JT gravity if a dilaton coupling is included \cite{Cao2021Mar}. The power laws associated with the continuous phase transition match those of the cSYK model. Given this, it is natural to ask how explicit one can make such analogies. For instance, would it be possible to have a gravitational model with the exact same thermodynamic potential and equation of state? Similar questions can be asked about the Lyapunov exponents reflecting the degree of chaos found in the respective models. 
   
   In this paper, we give partial answers to these questions. We explore the phase structure of deformed JT gravity and the cSYK model by comparing their partition functions. Our focus is on the on-shell physics, which corresponds to the solutions that minimize the action and characterize the leading order thermodynamics. To achieve this, we consider the $q/2$-body interacting complex SYK model. One can then derive the exact thermodynamic potential in powers of $1/q$. On the cSYK side, we place emphasis on the fluctuations away from on-shell, described by the Schwarzian \cite{Maldacena2016Nov}, by neglecting the $q$-dependent contributions. In the context of holography, the focus is usually placed on these off-shell fluctuations \cite{Davison2017,Gaikwad2020Feb}. Typically, these fluctuations cannot be ignored at nonzero temperatures. The resulting action can be expanded around the conformal solution to yield fluctuations described by the Schwarzian action. However, by expanding in $1/q$, we find that they are sub-leading, in orders of $1/q$, to the on-shell contributions \cite{Maldacena2016Nov}. The $1/q$ expansion, however, goes beyond this, also providing information about the harmonic oscillator-like phase, where conformal symmetry is strongly broken \cite{Azeyanagi2018Feb,Louw2022Feb}. This is because it provides the full phase diagram to leading order in $1/q$, hence it is not restricted to certain charge densities or low temperatures. 

	As for finding the candidate bulk dual, we start with a rather general model of deformed JT gravity. It is characterized by a dilaton potential energy $\Uu$ and a dilaton coupling $\Ww$ to Maxwell fields. In the context of the charged SYK model, such a theory has been proposed before as the low-energy dual \cite{Chowdhury2022}. Since the focus was placed on the low-energy limit, the considered deformations, were power laws. To capture the thermodynamics away from the strictly low-energy limit, we must consider more general deformations $\Ww$ and  $\Uu$. This is possible, since, like large-$q$ cSYK, the generally deformed model admits exact solutions \cite{Cai2020Aug}. Starting with some unknown potentials $\Ww$ and $\Uu$, we calculate various quantities. For instance, we find the general form of the equation of state (EOS), which is related to the Hawking temperature $T_{\text{H}}$, the Wald entropy $\Ss_{\text{W}}$ and the Arnowitt-Deser-Misner (ADM) mass $M$. We also find the associated Gibbs free energy $G$. All of these quantities are given in terms of the unknown functions $\Uu$ and $\Ww$, which we constrain such that we obtain the same thermodynamics as the cSYK model.
	
	By expressing the charge density as a function of the entropy, we are able to show that the same thermodynamic relations hold for both models. This relation allows us to identify the enthalpy on the SYK side, while the ADM mass corresponds to the enthalpy on the gravitational side. Equating these two enthalpies, we show that the equations of states also match. This requirement then fixes the potentials, identifying the sought deformation. It is further shown that their partition functions exactly match $Z_{\text{cSYK}} = Z_{\text{dJT}},$ in the regimes of interest. With this bulk dual, we go on to describe its gravitational properties, such as its scalar curvature, and how it relates to the condensed matter system. 

 We find two different dictionaries which still provide the same thermodynamics, These correspond to the two different analogies that one can draw between the van der Waals liquid, RN black holes, and the complex SYK model \cite{Kubiznak2012Jul,Louw2022Feb}.
	
    \section{The \texorpdfstring{$q$}{q}-dependent cSYK model\label{secSYKPt}} 
    We start from the $q/2$-body interacting cSYK model
	\cite{Fu2018}
	\begin{equation}
		\hat{\Hh} = J \sum_{\substack{1\le i_1< \cdots < i_{q/2}\le \Nn \\ 1\le j_1<\cdots< j_{q/2}\le\Nn}} X^{ i_{1}\cdots  i_{q/2}}_{j_{1}\cdots j_{q/2}} c^{\dag}_{ i_1} \cdots c^{\dag}_{ i_{\frac{q}{2}}} c_{j_{\frac{q}{2}}}^{\vphantom{\dag}} \cdots c_{j_1}^{\vphantom{\dag}}, \label{H}
	\end{equation}
	with a conserved U$(1)$ charge density $\hat{\Qq} = \frac{1}{\Nn}\sum_i c_i^\dag c_i - 1/2$, with expectation values $\Qq \in [-1/2,1/2]$, where $c^\dag, c$ are fermionic creation and annihilation operators, respectively. Here $\Nn$ is the number of lattice sites, hence the thermodynamic limit corresponds to taking $\Nn\to\infty$.  The couplings, $X$, are complex random variables with zero mean, and a variance $\overline{\,|X|^2} =  [q^{-1}(q/2) !]^2 [2/\Nn]^{q-1}$. We will work in the grand canonical ensemble 
    $$Z_{\text{cSYK}} = \text{tr}\,\exp(-\beta [\hat{\Hh} - \mu N (\hat{\Qq}-1/2)]).$$
	
    By considering $q/2$-body interactions instead of two-body interactions, one may solve the SYK model exactly, treating $1/q$ as an expansion parameter.  It was first pointed out by Davison et al. in \cite{Davison2017} that the equilibrium state described by $\Hh$ \eqref{H} tends to free fermions, for any non-zero charge density $\Qq = \Oo(q^0)$, in the large $q$ limit. This can be seen in the effective interaction strength
     \begin{equation}
         \Jj(\Qq) \equiv J [1-4\Qq^2]^{q/4-1/2} \label{effectiveInt}
     \end{equation}
     going to zero as $q\to \infty$. Even for small charge densities $\Jj(\Qq) \sim e^{-q \Qq^2} J$. To avoid this tendency, Davison et al. considered an altered  Hamiltonian $H_{\text{alt}}(\beta\mu)$, where the bare system coupling $J\to J_{\text{alt}}(\beta\mu)$ grows as a function of inverse temperature $\beta$ and chemical potential $\mu$ to compensate for the effective suppression.  The acquired $\beta\mu$-dependence of $H_{\text{alt}}(\beta\mu)$, however, leads to starkly different thermodynamics from $\Hh$ \eqref{H}, for any $q$ \cite{Ferrari2019Jul,Azeyanagi2018Feb}, as discussed in App. \ref{DavComp}. 

     \begin{figure}
	\includegraphics[width=0.9\linewidth]{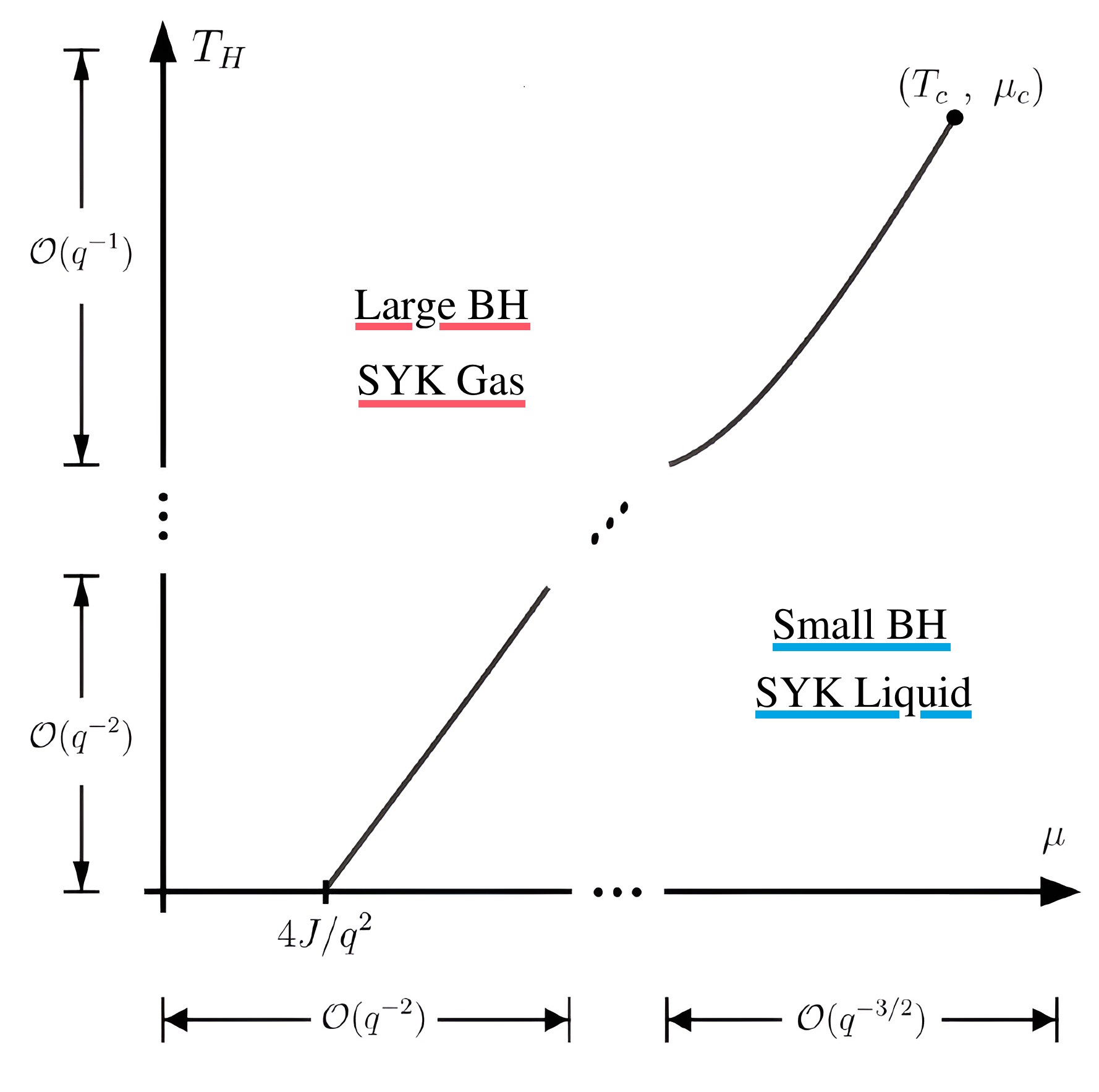}
    \caption{Schematic phase diagram for our particular deformed JT and cSYK models in different $q$-scaling regimes under the large-q limit condition. The upper regime encompasses the critical endpoint of the coexistence line. Close to the origin, there is a near-extremal phase transition.  \hspace*{\fill} \label{fig:tempptdiag}\label{HPSYKphasediag}}
    \end{figure}
     
     By not making any changes to the Hamiltonian \eqref{H}, we preserve the non-trivial thermodynamics at small fluctuations $\Qq = \Oo(q^{-1/2})$ away from $\Qq = 0$ \cite{Louw2023}. Remarkably, this unaltered cSYK model leads to a liquid-gas phase diagram which bares a striking resemblance to the small-large black hole phase diagrams found in black hole thermodynamics. The ``liquid" and ``gaseous" phases reflect their respective (charge) densities. In particular \eqref{H} exhibits a phase transition below a critical temperature $T_{\text{c}} = \Oo(q^{-1})$ or critical chemical potential $\mu_{\text{c}} = \Oo(q^{-3/2})$ \cite{Louw2023} because the temperature is $q$-dependent and the scaling transformation given in \cite{Davison2017} is broken.  Explicitly the critical point is at
     \begin{equation}
         T_{\text{c}} = 2 \Jj(\Qq_{\text{c}})/q, \quad \mu_{\text{c}} = 6 T_{\text{c}} \Qq_{\text{c}},\quad \Qq_{\text{c}} = \sqrt{3/(2q)}. \label{critPoint}
     \end{equation}

	Regarding the relation to gravity, there are two regimes of interest, the first considers a scaling $T = q^{-1} \tilde{T}$, $\mu = q^{-3/2} \tilde{\mu}$, with tilde'd quantities are $q$-independent. Around the transition point, the strongly coupled cSYK model dominates due to the relatively small charge densities. This rescaled regime corresponds to the IR regime, small $\beta\Jj$, hence both phases are maximally chaotic, reflected in their Lyapunov exponents saturating the Maldacena-Shenker-Standford (MSS) bound  $\lambda_L \to 2\pi T$ \cite{Maldacena2016}. This feature is shared with black holes. In particular, it is shared by both large and small black hole phases in the extended space \cite{Kubiznak2012Jul}. The corresponding phase transition also shares a universality class with that of the cSYK model. Both cases have mean-field critical exponents.

  \begin{figure}
    \label{muQ}
    \includegraphics[width=0.9\linewidth]{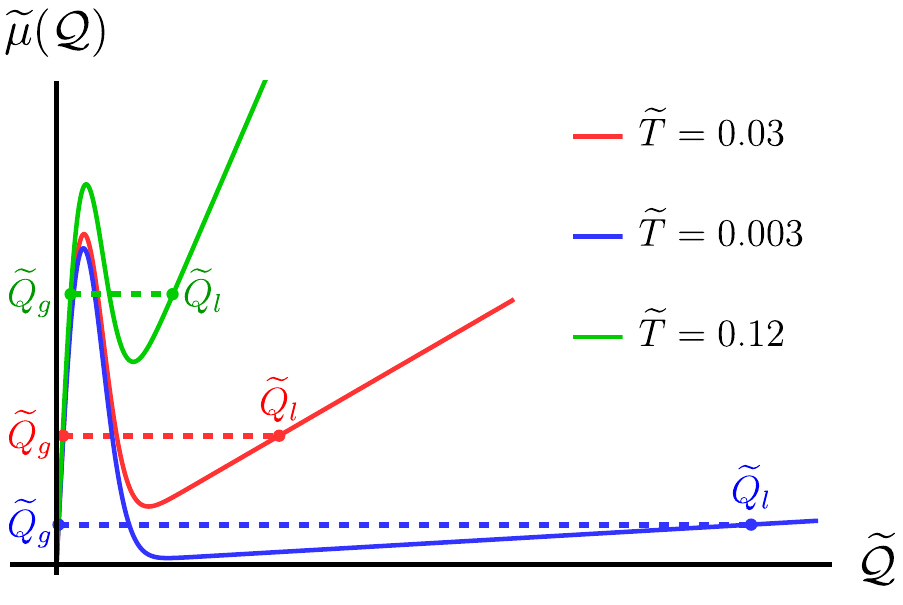}
    \caption{The chemical potential $\widetilde{\mu}$ as a function of the charge density $\widetilde{\mathcal{Q}}$, the dash lines of different colors represent the real physically acceptable solution which satisfies the Maxwell area law.}
    \end{figure}
	
    We further consider a second rescaled regime $T = q^{-2} \bar{T}$, $\mu = q^{-2} \bar{\mu}$, where barred quantities are held fixed as $q\to \infty$, with the corresponding phase diagram given in fig. \ref{HPSYKphasediag}. The gaseous phase in this regime corresponds to an uncharged, $\Qq = 1/q$ (in the large $q$ limit), and maximally chaotic SYK model. The liquid phase becomes incompressible and has an exponentially small entropy which tends to zero. The incomprehensibility stems from it reaching a maximal charge density, which is of the order $\Qq = \Oo(q^0)$. As noted before, such a large density fully suppresses the SYK interactions, yielding a free non-interacting model. The non-zero to zero entropy drop is analogous to the black hole to the thermal radiation Hawking-Page transition. In the large $q$ limit, the jump in charge density from $1/q$ to $1/2$, caused by a small perturbation $\mu_0 = 4 J/q^2$ to the chemical potential, is reminiscent of spontaneous symmetry breaking. To see how this behavior of the charge density nearby the coexistence line emerges, we can go back to the first scaling regime and plot the chemical potential $\widetilde{\mu}$ as a function of charge density $\widetilde{\Qq}$. One can directly find that as the decreasing of the temperature $\widetilde{T}$, the charge density of liquid phase $\widetilde{\mathcal{Q}}_l$ goes to infinity which implies that the corresponding non-rescaled charge density $\mathcal{Q}_l$ is the order of $\mathcal{O}(q^0)$ as the rescaled temperature $\widetilde{T}$ goes to zero. At the same time, the charge density of the gas phase vanishes. This phenomenon indicates the jump in the charge density we described above. This highlights a difference between the two rescaled regimes. In the first regime, close to the critical point, the liquid and gaseous charge densities are of the same order. As such, for the specific rescaled quantities $(\tilde{\mu},\tilde{T},\tilde{\Qq})$, the limit as $q\to \infty$ is well-defined.

    In the regime associated with the zero $T$ limit, we have two different scalings in the charge density, namely $\Qq = \Oo(1/q)$ and $\Qq = \Oo(q^0)$. As such, the two charge densities diverge from one another. For any finite, but large $q$, the phase transition still exists. In the limit $q \to \infty$, one can, however, argue that this phase transition no longer makes sense due to this diverging separation. Similarly, the limit of a spherical to a flat Euclidean space as the parameter $k\to 0$ \cite{Chamblin1999}, the phase transition also disappears, in which the parameter $k$ represents the topological parameter of the RN-AdS black hole \cite{ammon_erdmenger_2015}. In this sense, one might be able to associate $k$ with $1/q$.

    \section{The deformed JT gravity model \label{secEMdgrav}}
	
    We consider general deformed JT gravity \cite{Witten2020Jun} together with coupling to a  Maxwell field \cite{Cai2020Aug}, with action 
	$$I[\varphi,A] = G_{\text{N}}^{-1} \nint[M]{^2 x} \sqrt{-g} \Ll(\varphi, A) + I_{\text{bdy}}.$$
	Here the boundary action contribution $I_{\text{bdy}}$, described in App. \ref{AppFreeEnergy}, regularizes the theory.  In $(1+1)$-dimensions, the constant $G_{\text{N}}$ is dimensionless in natural units $\hbar = c = 1$. Its inverse will play the role of the large parameter $N$ selecting out the on-shell solution in the classical limit. To have a well-defined limit, we must focus on ''intensive" quantities, for instance focusing on the intensive bulk Lagrangian density
	\begin{equation}
		\Ll(\varphi, A) = \frac{\varphi}{4\pi} \Rr_2 +P\, \Uu(\varphi) - \frac{\Ww(\varphi)}{4}  F(A)^2, \label{BulkLag}
	\end{equation}
	instead of $\Ll/G_{\text{N}}$. Here $F_{\mu\nu} = \p_{\mu} A_{\nu}-\p_{\nu} A_{\mu}$ is the electromagnetic tensor and $\Rr_2$ is the 2-dimensional Ricci scalar. The dilaton $\varphi$ couples to the electromagnetic field via a term $\Ww(\varphi)$. The field also has its own potential energy  $P\Uu(\varphi)$, where we have a thermodynamic pressure term $P$. This pressure is associated with a negative cosmological constant \cite{Kubiznak2017Feb}, which is the pressure of empty space. Since the characteristic length scale is associated with the scalar curvature at the conformal boundary, we would assume it to be related in some way to the interacting contribution of the quantum system.
		
    We assume the black hole solution, in the Schwarzschild gauge, takes the form
	$ds^2 = - f(r) dt^2 + dr^2/f(r)$. Solving the Euler-Lagrange equations, see App. \ref{secEMDaction},  we find the dilaton field solution \eqref{dilSol} $\varphi = \gamma r$, where $\gamma$ is the dilaton coupling strength. Setting $\gamma=1$ amounts to measuring distance in units of $\gamma$. We also have the emblackening factor \eqref{f}
	\begin{equation}
		f(r)/(4\pi) = -  M + P V(r) - Q_{\text{B}} A_t(r)/2, \label{f(r)}
	\end{equation}
	for some integration constant $M$ and black hole charge $Q_{\text{B}}$. Here we have defined the anti-derivatives
	\begin{equation}
		V(r)=  \nint{r}\, \Uu(r), \quad A_t(r)= Q_{\text{B}}\nint{r} \frac{1}{ \Ww(r)}. \label{VPhifunc}
	\end{equation}
	
	By definition, the event horizon is at the root $r=r_{\text{H}}$ of $f(r)$, i.e., $f(r_{\text{H}})=0$.  With this, \eqref{f(r)} implies
	\begin{equation}
		M = P V_{\text{th}} + Q_{\text{B}} \Phi_{\text{th}}/2, \quad V_{\text{th}} \equiv V(r_{\text{H}}),\;\Phi_{\text{th}} \equiv -A_t(r_{\text{H}}), \label{ADMMass}
	\end{equation}
	where we have defined the thermodynamic quantities as \eqref{VPhifunc} evaluated at the horizon $\varphi_0 =  r_{\text{H}}$. For instance, in black hole chemistry, the pressure is conjugate to the volume \cite{Dolan2016May} leading to the identification of  $V_{\text{th}}$ as the thermodynamic volume. We identify, as usual, the Hawking temperature as $T_{\text{H}} \equiv f'(r_{\text{H}})/(4\pi)$ which is the conjugate to the Wald entropy \cite{Wald1993Oct} $\Ss_{\text{W}} = r_{\text{H}}$.  As such the function $M(\Ss,P,Q_{\text{B}})$  satisfying the differential relation
	\begin{equation}
		d M =  \Phi_{\text{th}} d Q_{\text{B}} + V_{\text{th}} dP+T_{\text{H}} d\Ss_{\text{W}}, \label{ADMdifrel}
	\end{equation}
    which is the first law of (black hole) thermodynamics \cite{Kubiznak2017Feb}, which also serves to define the thermodynamic volume. Indeed, it can be identified as the ADM mass \cite{Gegenberg1995}. In considering $\Phi_{\text{th}}$ to be the black hole's chemical potential \cite{Cai2020Aug}, we may also view it as an enthalpy. From \eqref{ADMdifrel},  using $\Ss_{\text{W}} = r_{\text{H}}$, we may also obtain the EOS
	\begin{equation}
		T_{\text{H}} = \left(\frac{\p M}{\p \Ss_{\text{W}}}\right)_{P,Q_{\text{B}}} = P V'(r_{\text{H}}) - \frac{Q_{\text{B}}}{2} A_t'(r_{\text{H}}), \label{EOSJT}
	\end{equation}
	where, unless specified otherwise, derivatives are evaluated keeping $P$ and $Q_{\text{B}}$ constant, $V'(r_{\text{H}}) \equiv (\p_{r_{\text{H}}} V(r_{\text{H}}))_{P,Q_{\text{B}}}$.
	
	The thermodynamic potential which selects out the favorable state is the Gibbs free energy \cite{Kubiznak2017Feb} $G(T_{\text{H}},P,Q_{\text{B}}) = M - T_{\text{H}}\Ss_{\text{W}}$. 
	This is identified with the on-shell action \eqref{Gibbs} of the uncharged black hole dual to the described charged system. All other expressions would remain unchanged if we had instead worked with this uncharged dual from the start. The Gibbs free energy also arises naturally in the dimensionality reduction of $(3+1)$-dimensional charged black holes \cite{Cai2020Aug}. 

\section{Matching the partition functions \label{secGrSYKrel}}
    Our goal is to find the gravitational Lagrangian dual to the cSYK model, which is defined by the yet to be determined potentials $\Uu$,$\Ww$. Equivalently, we may focus on the related anti-derivatives $V$, $A_t$ defined in \eqref{VPhifunc}. We do this by focusing on the large $q$ cSYK model's grand potential
	\begin{equation}
		\Omega \equiv -T\ln Z_{\text{cSYK}} /N = E + (1/2-\Qq) \mu - T \Ss, \label{grandPot}
	\end{equation} 
	with the interaction energy \cite{Louw2022Feb} 
	\begin{eqnarray}
		E \sim  -2\epsilon(\Qq)/q^{2}, \quad \epsilon(\Qq) \equiv \Jj(\Qq) \sin(\pi v/2) \label{IntEn}
	\end{eqnarray} 
	and entropy density $\Ss = \Ss_{2}(\Qq)-(\pi v/q)^2/2$, as shown in App. \ref{twoStateEntropy}, where 
	\begin{equation}
		\Ss_{2}(x) \equiv - \frac{1-2 x}{2} \ln\!\left\vert\frac{1-2 x}{2}\right\vert- \frac{1+2 x }{2} \ln\!\left\vert\frac{1+2x}{2}\right\vert, \label{entExpl}
	\end{equation}
	which is an even function of $x$. Here $v$ is the solution to the closure relation $\Jj(\Qq)/T = \pi v \sec(\pi v/2) $ \cite{Maldacena2016Nov}, which is also related to the Lyapunov exponent as $\lambda_L = 2\pi T v$.

	The phase transition is reflected in the EOS
	\cite[eq.(43)]{Louw2022Feb}
	\begin{equation}
		T  = \frac{\mu - 4\Qq \epsilon/q}{2\tanh^{-1}(2\Qq)}, \label{TSYKEOS}
	\end{equation}
	becoming three-to-one for $T<T_{\text{c}}$, or $\mu<\mu_{\text{c}}$, where the critical temperature and critical chemical potential scales as $T_{\text{c}} = \mathcal{O}(1/q)$ and $\mu_{\text{c}} =\mathcal{O}(q^{-3/2})$.  Equation (\ref{TSYKEOS}) is q-dependent since for example, it breaks the  scaling symmetry $T\rightarrow T/q^2$, $\mu \rightarrow \mu/q^2$, and $\mathcal{Q}\rightarrow \mathcal{Q}/q$.  Note that this equation is invalid for $\Qq=0$, amounting to division by zero, in which case the temperature becomes an independent free parameter. One may show that this EOS remains valid for large $q$, for any polynomial (in $q$) scaling for temperature and chemical potential \cite{Louw2023}, i.e., the cases we consider.

	To have matching thermodynamics, we not only require the same thermodynamic potentials $\Omega,G$, but also matching  equations of states. If the quantity $\Omega +T\Ss$:
	\begin{equation}
	H \sim  -2\epsilon(\Qq)/q^{2}+(1/2-\Qq) \mu \label{enthalpy}
	\end{equation}
	satisfies the same relation as the mass \eqref{EOSJT}
	\begin{equation}
		T = \left(\frac{\p H}{\p \Ss}\right)_{\mu,J} = \left(\frac{\p \Qq}{\p \Ss}\right)_{\mu,J} \left(\frac{\p H}{\p \Qq}\right)_{\mu,J}, \label{enthalpyEntropy}
	\end{equation}
	i.e., yields the same EOS, then it can also be identified with the enthalpy. The above relation may be rewritten as
	\begin{equation}
		\beta\left(\frac{\p H}{\p \Qq}\right)_{\mu,J}= \left(\frac{\p \Ss}{\p \Qq}\right)_{\mu,J} = -2 \tanh^{-1}(2\Qq) - \p_\Qq \frac{(\pi v/q)^2}{2} \label{dSOdQ}
	\end{equation}
	where unless specified otherwise, we assume that $\mu,J$ are kept constant, meaning that $	v'(\Qq) \equiv	\left(\p_{\Qq} v(\Qq)\right)_{\mu,J}$.  	
	Using the closure relation $\beta\Jj(\Qq) \cos(\pi v/2) = \pi v$, the left-hand-side of \eqref{dSOdQ} reduces to
$$ -2\beta\epsilon'(\Qq)/q^2-\beta\mu =4 \Qq \beta\epsilon(\Qq)/q-\pi^2 v'(\Qq)v/q^2-\beta\mu.$$
 Finally, from \eqref{TSYKEOS}, we have $4\Qq \beta\epsilon/q = \beta\mu-2\tanh^{-1}(2\Qq) $, which leaves the right-hand side of \eqref{dSOdQ}, thus finished the proof identifying $H$ as an enthalpy. Considering \eqref{EOSJT} and \eqref{enthalpyEntropy} we note that the same equation of state is obtained if we identify the temperatures and entropies and enthalpies with another which also then implies that $G = \Omega$, since 	$G = M - T_{\text{H}} \Ss_W$ and $\Omega = H - T \Ss$. This (partial) dictionary is summarized in table \ref{Dictionary}. With these identifications, one finds not only an isomorphism between the EOS's and thermodynamic potentials, but \emph{equivalent} partition functions $$Z_{\text{dJT}} = e^{-\beta N G} = Z_{\text{cSYK}} = e^{-\beta N\Omega}.$$
	Since the thermodynamics is uniquely encoded by the partition function and EOS, we also have the exact phase diagram matching Fig. \ref{HPSYKphasediag}. The same holds true in the maximally chaotic regime, where the phase diagram has been given in \cite{Louw2023}.
 
	\begin{table}
		\centering
		\caption{Dictionary between the thermodynamics of the q-dependent cSYK model and deformed JT (dJT) gravity. Each row identifies the two quantities which equate to another.\hspace*{\fill} \label{Dictionary}}  
		\begin{tabular}{|c|c | c  |} 
			\hline
			Model& {cSYK} & dJT \\
			\hline
			large parameter& $N$ & $1/G_{\text{N}}$\\
			enthalpy & $H$ \eqref{EOSJT}& $M$ \eqref{ADMMass}\\
			entropy density & $\Ss$ & $\Ss_{\text{W}}$\\
			temperature & $T$ \eqref{TSYKEOS} & $T_{\text{H}}$ \eqref{EOSJT}\\
			thermodynamic potential & $\Omega$ \eqref{grandPot}& $G$\\
						\hline
		\end{tabular}
	\end{table}
 
  To further specify the dictionary, we consider the differential relations of the two models. For the cSYK model, we have
\begin{equation}
\left(\frac{\p \Omega}{\p J}\right)_{\mu,T} = \frac{E}{J},\quad	\left(\frac{\p \Omega}{\p \mu}\right)_{J,T} = \frac{1}{2}-\Qq \label{dOmega}
\end{equation} 
while for the gravitational model's Gibbs free energy, we have
\begin{equation}
\left(\frac{\p G}{\p Q_{\text{B}}^2}\right)_{P,T} =  \frac{\Phi_{\text{th}}}{2 Q_{\text{B}}},\quad \left(\frac{\p G}{\p P}\right)_{Q_{B},T} = V_{\text{th}}.\label{GibbsGr}
\end{equation}
By comparing these two, we note the two possible options given in Table. \ref{Dictionary2}.

\begin{table}
    \caption{Dictionary relations for parameter and conjugate pairs   \hspace*{\fill} \label{Dictionary2}
    }
    \begin{tabular}{|c || c | c |} 
    \hline
    cSYK & dJT \textbf{(a)} & dJT \textbf{(b)}\\
    \hline
    $\mu$, $\;1/2-\Qq$& $Q_{\text{B}}^2$, $\;\frac{\Phi_{\text{th}}}{2 Q_{\text{B}}}$ & $P$, $\;V$\\
    $J$, $\;E/J \quad$ & $P$, $\;V$ & $Q_{\text{B}}^2$, $\;\frac{\Phi_{\text{th}}}{2 Q_{\text{B}}}$\\
    \hline
    \end{tabular}
\end{table}
This choice will \emph{not} influence the thermodynamics, except for its interpretation on the black hole side. Such that we do not restrict ourselves to a particular choice, we typically use the notation on the condensed matter side.  These two options in fact directly overlap with the two different analogies which can be drawn between the van der Waals liquid, RN black holes, and the charged SYK model \cite{Kubiznak2012Jul,Louw2022Feb}.

\subsection{Equivalence of thermodynamics}

\begin{figure}
		\centering
		\includegraphics[width=0.8\linewidth]{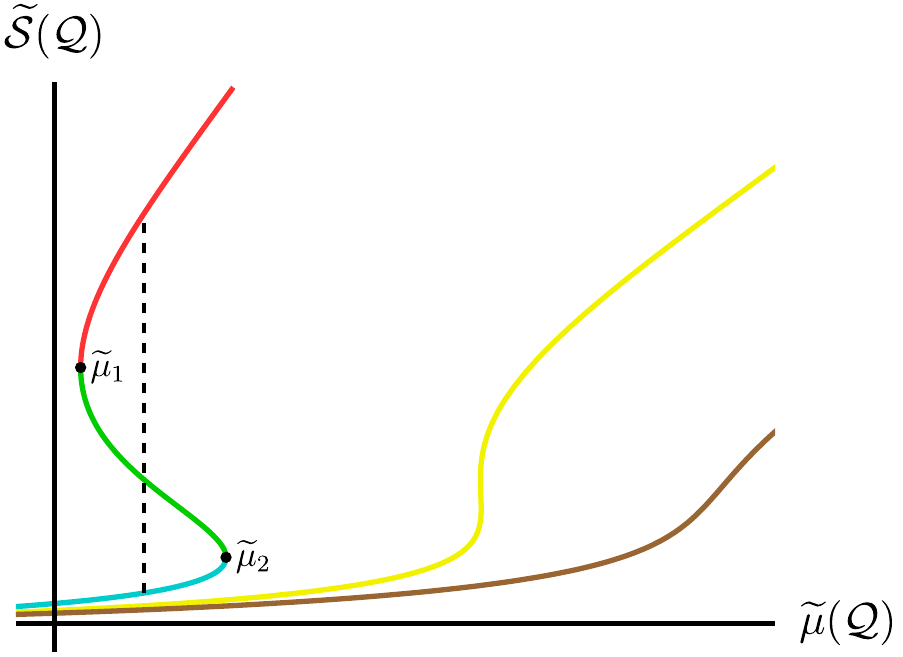}
		\caption{The red, green, and blue curves represent the three phases of black hole, small, medium, and large, respectively. The dashed line stands for the thermodynamically favorable solution, in which the area of both sides is the same (Maxwell area law). The yellow line and brown lines are for $\widetilde{T}=\widetilde{T}_{\rm crit}$ and $\widetilde{T}>\widetilde{T}_{\rm crit}$ respectively.}
		\label{S-mu}
    \end{figure}

Since we know the thermodynamics match, we can consider the equation of state \eqref{TSYKEOS} in the context of the black hole's thermodynamics. Below a critical chemical potential, associated with either  charge (dictionary \ref{Dictionary2}.a) or pressure (dictionary \ref{Dictionary2}.b), or temperature, \eqref{TSYKEOS} becomes three-to-one. The Wald entropy is equal to the horizon radius, but also equal to the cSYK entropy $r_{\text{H}}=\Ss(\Qq)$, given the dictionary table. \ref{Dictionary}. As such the three different charge densities $\Qq$ correspond to three different entropies (horizon radii), i.e., three different states as plotted in fig. \ref{S-mu}. These entropies $$\Ss \in \{\Ss_{\text{large BH}}, \Ss_{\text{unstable BH}}, \Ss_{\text{small BH}}\},$$ correspond to three different horizon radii; hence we have a large black hole, a small unstable black hole and a small stable black hole, as expected from a charged extended space system \cite{Kubiznak2012Jul}. Those with a positive specific heat, corresponding to negative horizon curvature, are the stable phases \cite{Witten2020Jun}. 

These entropies exactly correspond to the phases of the gaseous, unstable liquid, and stable liquid phases of the cSYK model, reflected in the different charge densities $\Qq$. These three phases are seen in the three-to-one behavior in the rescaled chemical potential $\widetilde{\mu}(\Qq)$ (or temperature) as a function of entropy $\widetilde{\Ss}(\Qq)$. Between the temperature $\widetilde{\mu}_1 \sim \widetilde{\mu}_2$, there are three different phases corresponding to the three different horizon radii. The thermodynamically preferred radius corresponds to the minimum Gibbs free energy between the three.
 
    It is important to note that since the partition functions and equations of states exactly overlap, given the dictionary \ref{Dictionary}, we are guaranteed to have equivalent thermodynamics for both dual models. This means that they share the same critical exponents, given in table \ref{exponents}, hence the same universality class. Here we take a moment to describe the thermodynamics from the gravitation perspective. This is done by translating the known results for the $q$ body cSYK model \cite{Louw2023} into gravitational language via the dictionary.

	\begin{table}[h]
		\caption{Tables of critical (left) and effective (right) exponents \label{exponents}}
		\begin{minipage}[c]{0.4\columnwidth}
		\begin{tabular}{|c c c c |} 
			\hline
			$\alpha$  & $\upbeta$ & $\gamma$ &  $\delta$\\
			\hline
			$0$ & $1/2$ & $1$  & $3$\\ 
			\hline
		\end{tabular}
		\end{minipage}
		\begin{minipage}[c]{0.48\columnwidth}
			\begin{tabular}{|c c c | c c|} 
				\hline
				$\alpha_\mu$ & $\upbeta_\mu$ & $\gamma_\mu$ & $\alpha_\Qq$ & $\gamma_\Qq$\\
				\hline
				$2/3$ & $1/3$ & $2/3$ & $0$  & $1$\\
				\hline
			\end{tabular}
		\end{minipage}
	\end{table}

To get some idea of the interpretations on the gravitational side, let us for the moment consider Table \ref{Dictionary2}.b. The charge density is provided in color on these diagrams and is then directly related to the thermodynamic volume of the black hole $V_{\text{th}}= 1/2- \Qq$. Note that when we approach the boundary, we consider smaller values of $\Qq$, corresponding to larger volumes.

Various power laws emerge as the critical point $(P_{\text{c}},T_{\text{c}})$ is reached which can differ from the critical exponents. This is due to a feature well known in the field of statistical mechanics known as field mixing \cite{Wang2007May}. The prototypical example is that of the van der Waals liquid. These \emph{effective} power laws are still physically relevant. For instance, the specific heat will diverge as $C_{P} \propto |T-T_{\text{c}}|^{-2/3}$, i.e. $\alpha_P = 2/3$. Given its relation to the Ricci scalar \eqref{RrCp}, we note that this ensures a finite horizon curvature.	It remains well-defined at constant volume $C_V \propto T \sim t^0$, as is common to RN system \cite{Zaanen2015Nov}.	The remaining \emph{effective} exponents can be obtained from \cite{Louw2023}, and are listed in \ref{exponents}.

The equivalence is over the entire coexistence line, meaning that we have the same thermodynamics also in the regime where the quantum model has a chaotic-to-nonchaotic transition $T = \Oo(q^{-2})$, $P = \Oo(q^{-2}) $. Here the chaotic phase corresponds to the maximally large black hole $r_{\text{H}}= r_{\text{max}}$. The nonchaotic phase on the quantum side corresponds to an evaporated black hole, where the horizon radius goes to zero $r_H\to0$. This transition occurs at a pressure $P_0 = 4Q_{\text{B}}/q^2$. We will further consider the degree of chaos, a dynamical property, in the next section.

\section{Metric dual to the cSYK model \label{gravAsp}}

To make the mapping more explicit, we must fully specify the functions $\Uu$ and $\Ww$ which define the dJT model. 	Extending the identification $r_{\text{H}}=\Ss(\Qq)$ to all radii, we have the equation $r=\Ss(x)$, or the inverse $x(r) = \Ss^{-1}(r)$. When evaluated at the horizon, we find the order parameter $\Qq = x(r = r_{\text{H}})$. We perform this inversion in various regimes in  App.\ref{IntSYKrel}. Given the above, we can fully specify the functions $V$ and $A_t$, hence $\Uu$ and $\Ww$, given \eqref{VPhifunc}. In other words, we can fully specify the particular deformation. Using the relations in \eqref{VPhifunc}, we have that 
$$\Qq'(r_{\text{H}}) = \begin{cases}
    -2/\Ww(r_{\text{H}}) & (a)\\
    -\Uu(r_{\text{H}}) & (b)
\end{cases} $$
which is equal to $1/\Ss'(\Qq)$. With this, we note that $\Ss'(\Qq)$ measures the coupling to the Maxwell fields given dictionary $(a)$, while dictionary $(b)$ yields the reciprocal dilaton potential. Hence, $(a)$ identifies the U$(1)$ charges on the gravitational side with that of the condensed matter side. Using either of the tables \ref{Dictionary2}.a or \ref{Dictionary2}.b would yield the enthalpy ``functions''
	\begin{equation}
		P V(r) - Q_{\text{B}} A_t(r)/2 = \mu [1/2-x(r)] -2\epsilon(x(r))/q^2. \label{f(x)part}
	\end{equation}  
Here the second term stems from the relation with the interaction energy density function \eqref{IntEn}
	\begin{equation}
		\epsilon(x) \equiv \Jj(x) \sin(\pi v(x)/2),\quad \Jj(x)\sim [1-4x^2]^{q/4} J. \label{Phifunc}
	\end{equation}
		Using \eqref{f(x)part} we find the metric corresponding to the cSYK model, defined by the emblackening factor \eqref{f(r)} written directly in terms of the dual condensed matter model's parameters
	\begin{equation}
		f(r)/(4\pi) = \mu [\Qq - x(r)] + 2\epsilon(\Qq)/q^2-2\epsilon(x(r))/q^2.\label{fmet}
	\end{equation}
The roots of this function yield the horizons. The largest root is the event horizon $r_{\text{H}}$ of the black hole, i.e., $x(r_{\text{H}})=\Qq$. The smaller root $r_-$, corresponding to large $x$, is the Cauchy horizon. For instance, where the interaction energy becomes exponentially small $\epsilon(x) \sim e^{-q x^2}$ we have a root at $x(r_-) = \Qq+2\epsilon(\Qq)/(\mu q^2)$.

	As before, the temperature is obtained from the function $f'(r_{\text{H}})$. For other values of $r$, we define the function
	\begin{equation}
		\Tt(x) \equiv	\frac{f'(r(x))}{4\pi} 	=  \frac{\mu - 4 x \epsilon(x)/q}{2\tanh^{-1}(2 x)} \label{f'(r)}
	\end{equation}
	where $T = \Tt(\Qq)$. With this, the closure relation becomes 
	\begin{equation}
	\Jj(x)/\Tt(x) =  \pi v(x)\sec(\pi v(x)/2)\label{clos}.
	\end{equation}
        Solving \eqref{clos} in the limiting cases, we find
	\begin{equation}
		\frac{\epsilon(x)}{\Jj(x)} \sim \begin{cases} 1 + \Oo(\Tt^2(x)/\Jj^2(x)),\\  \frac{\Jj(x)}{\mu}\,\tanh^{-1}(2 x), & \text{for } x = \Oo(q^{0})\end{cases} \label{PhiV}.
	\end{equation}
	Evaluating \eqref{f'(r)} at the horizon, where $x(r_{\text{H}}) = \Qq$, yields the cSYK EOS \eqref{TSYKEOS} as expected from a dual theory. 

  Evaluated at the horizon, the curvature may be written as
	\begin{equation}
		\Rr_2(r_{\text{H}}) = -4\pi  \left(\frac{\p T_{\text{H}}}{\p \Ss_{W}}\right)_{\mu,J} = -\frac{4\pi  T_{\text{H}}}{C_\mu}, \label{RrCp}
	\end{equation}
	where $C_\mu$ is the heat capacity at constant chemical potential
	\begin{equation}
		C_\mu \equiv T_{\text{H}} \left(\frac{\p \Ss_W}{\p T_{\text{H}}}\right)_{\mu,J}. \label{Cp}
	\end{equation}

 For $\Qq = \Oo(q^0)$, the SYK interactions are suppressed, yielding a near-free system $\Qq \sim \tanh(\beta\mu/2)/2$ with specific heat $C_\mu \sim 2(\beta \mu)^2 e^{-\beta\mu}$ as entropy tends to zero ($\Qq \to 1/2$). This means that the curvature at the horizon blows up as the dual system becomes a free Fermi gas, as was found in \cite{Ruppeiner2018Jun}. In this sense, the mapping is a weak-strong duality. An analogy would be how the shear viscosity diverges in the free theories with holographic duals considered in \cite{Policastro2001,Son2007}.	

Given the above discussion, we can now gain an idea of the metric dual to the cSYK model. Recall that the stable phases have positive specific heat. Noting that $f^{(1)}(r_{\text{H}}) = 4\pi T_{\text{H}}$ and $f^{(2)}(r_{\text{H}}) = 4\pi T_{\text{H}}/C_{\mu}$, we may express the near-horizon emblackening factor as
\begin{equation}
	f(r_{\text{H}}+\delta) = 4\pi T_{\text{H}}\delta\left(1+\delta/(2C_\mu)\right)   + \Oo(\delta^{3}). \label{NHf}
\end{equation}
As such the stable phases, above and near the horizon, will have a positive concave-up emblackening factor.

	\subsection{Need for an IR cutoff \label{IRcut}}
	
	As $x \to 1/2$, the interaction energy contributions are fully suppressed, leaving a free theory. As such, we need only invert $\Ss_2(x)$, which yields
	\begin{equation}
		x(r) \sim \frac{1}{2} - \frac{r}{\ln(1/r)} \quad\xrightarrow{r\to 0^+}\quad 1/2.
	\end{equation}
	We have $r_{\text{H}} = 0$ corresponding to $\Qq =1/2$. We, however, exclude this point from our space, i.e., $r>0$. From this, we also note that $r\ge r_{\text{H}}$, i.e., when $x(r)\le \Qq$. A naive calculation of $x$, when $x$ is small, yields the inverse $x(r) \sim \sqrt{(\ln2-r)/2}$. The diverging second derivative at $r=\Ss(0)$ would also yield a diverging scalar curvature $\Rr_2(r) = -f^{(2)}(r)$. This simple expression is due to the simplicity of the two-dimensional static metric we have, yielding simple Christoffel symbols.

 As mentioned in Sec. \ref{secGrSYKrel}, the EOS \eqref{TSYKEOS} is not valid for $\Qq=0$, seen in its diverging temperature. This is because, on the condensed matter side, it determines the chemical potential 
 $$\mu = 2 T \tanh(2\Qq) + 4 \Qq \epsilon/q$$
 rather than the temperature. As such, $\Qq=0$, directly implies $\mu=0$, leaving $T$ a free variable. In the form of \eqref{TSYKEOS}, we are thus effectively dividing by zero, when $\Qq=0$. Since this corresponds to a zero charge SYK model, this point $r=\ln 2$ is also where the EOS \eqref{TSYKEOS} fails. We fix this by limiting our scope to small but non-zero charge densities. On the gravitational side, this means that we consider a minimal $x=x_{\text{min}}\neq 0$. This is equivalent to introducing an IR cutoff radius $r_{\text{max}}$. The square root is then modified to
	\begin{equation}
		x(r) \sim \sqrt{x_{\text{min}}^2 + (r_{\text{max}}-r)/2 }\quad\xrightarrow{r\to r_{\text{max}}}\quad x_{\text{min}}  \label{xcutoff2}
	\end{equation}
 As such, to have a well-defined theory, we should have some non-zero minimum value $\Qq = x_{\text{min}}$. Such a minimum appears when considering a particular IR cutoff $r_{\text{max}}$. We choose this cutoff such that our theory will satisfy two conditions:
	
\emph{(I)}	Given the cutoff we have access to the full liquid-gas coexistence line of the SYK model. 

\emph{(II)}	The scalar curvature $\Rr_2(r_{\text{max}})$ remains finite for any finite value of $q$. This condition would, for instance, be violated given an emblackening factor $f(r) \propto \sqrt{r_{\text{max}}-r}$, which has both a diverging temperature function (related to $f'(r)$) and scalar curvature (related to $f^{(2)}(r)$).

One choice in cutoff is such that we include the minimum charge density which occurs along the coexistence line in the cSYK model, $x_{\text{min}} = 1/q$ \cite{Louw2023}. Given that the entropy function relates $x$ to the radius, we substitute this value to find 
\begin{equation}
	r_{\text{max}} = \ln 2 - \frac{4 +\pi^2+ \Oo(q^{-2})}{2q^2} .
\end{equation}
Note that for both the first or second rescaled regimes
\begin{equation}
    \mu = q^{-3/2}\tilde{\mu} = \Oo(q^{-3/2}),\quad \mu = q^{-2} \bar{\mu} = \Oo(q^{-2})
\end{equation}
we are guaranteed a small temperature function at the cutoff
\begin{equation}
	\Tt(x_{\text{min}}=1/q) \sim  q \mu/4- J/q +\Oo(q^{-3/2}).
\end{equation}  
Further motivations for this choice are provided in App.\ref{detailsIRCutoffs}. From the above, we also note the endpoint of the coexistence line $\mu_0 = 4 J/q^2$, corresponding to zero temperature. At the boundary $\Tt(x_{\text{min}})$, \eqref{Cp} is given by $\beta C_\mu \sim 16/\mu$, for $\mu$ of order $q^{-3/2}$ or lower. Now using \eqref{Cp}, we find the scalar curvature $\Rr_2(r_{\text{H}}) =	-4\pi T/ C_{\mu}$, yielding the boundary curvature $\Rr_2(r_{\text{max}}) =	-\pi q^{3}\mu/4$
which is indeed finite for any finite $q$, hence our chosen cutoff satisfies condition \emph{(II)}. The dictionary \ref{Dictionary2}.b is required if we  wish to identify the pressure with the cosmological constant, as standard in black hole chemistry \cite{Kubiznak2017Feb}. A different cutoff would yield a different curvature. Since the cutoff is not unique, one could view this specific cutoff as being the most appropriate in that it yields the expected curvature.

From the above, we note that the near-extremal limit, we are left with $f(r) \propto -\pi q J (r-r_{\text{H}})^2$.  Close to the cutoff, for small $\delta = q^2 (r-r_{\text{max}})$, we have \eqref{nearCutfp} $f'(r) = q \pi \mu (1-\delta/2)^{-1/2} - q\pi \mu_0$, implying the emblackening factor
\begin{align}
f(r)  &= f(r_{\text{max}}) + q \pi [\mu-\mu_0]\delta/2 + \frac{q \pi \mu }{8} \delta^2   +\Oo(\delta^3)
\end{align}
working to explicit order $\Oo(q^{-1})$.

There are multiple other choices of cutoffs that would satisfy both above conditions. One could also consider UV cutoffs to regularize the theory at smaller distances. 

\section{Comparison of Lyapunov exponents \label{ChaosTraj}}
	
	In this section, we wish to compare the dynamical properties of the two models with matching thermodynamics. While it was true that the choice of particular dictionary in Table \ref{Dictionary2} did not affect the thermodynamics, the same cannot be said about the dynamics. This is because we are choosing which cSYK term should be identified with the electrical field. Here we will consider both cases. We focus on their Lyapunov exponents measuring the sensitivity to initial conditions. We write the Lyapunov exponent as $\lambda_L = 2\pi v T$. For the SYK model, $v$ is the solution to the closure relation \eqref{clos} $\beta \Jj(\Qq) = \pi v \sec(\pi v/2)$. In the maximally chaotic regime $T = q^{-1} \tilde{T}$, $\mu = q^{-3/2} \tilde{\mu}$ with tilde'd quantities are $q$-independent, it is solved by	
	\begin{equation}
		v = 1 -2 q^{-1} \tilde{T}/\Jj(\Qq) +\Oo(q^{-2}) \xrightarrow{q\to \infty} 1.\label{vChaotic}
	\end{equation} 
 
    The liquid phase becomes near-integrable in the second rescaled regime $\beta = q^{2} \bar{\beta}$, $\mu = q^{-2} \bar{\mu}$, where barred quantities are held fixed as $q\to \infty$. In this same regime, the gaseous phase remains maximally chaotic. The tendency to integrability is driven by its large charge density $\Qq = \Oo(q^0)$ which suppresses the effective coupling $\Jj(\Qq) \sim J e^{- q\Qq^2}$, leading to an exponentially small Lyapunov exponent $v =  q^2\bar{\beta}\Jj(\Qq)/\pi \xrightarrow{q \to \infty} 0$.

	For a non-extremal black hole, the maximal Lyapunov exponent is usually given by the surface gravity $\kappa = f'(r_{\text{H}})/2 = 2\pi T_{\text{H}}$ \cite{Hashimoto2017Jan} which is MSS bound \cite{Maldacena2016}. We find $\lambda_{L}$ by focusing on the near-horizon trajectory of a charged particle close to the black hole.  The corresponding equations of motion are \cite{Lei2022Apr} $\dot{r} = \pi_r f$, $\dot{t} = -[\pi_t + Q_{\text{e}} A_t]/f$ and
	\begin{equation}
		\dot{\pi}_r = - \pi_r^2/(2f') - \dot{t}^2 f'/2-Q_{\text{e}} A_t' \dot{t},
	\end{equation}
	where $\pi_t$ and $\pi_r$ are the $t$ and $r$ components of particle momentum, respectively. The particle's charge is given by $Q_{\text{e}}$ and $A_t=-\Phi$. Note that we are focusing on the particle's geodesic for a non-dynamic metric. As such, an implicit assumption is the particle's back-reaction on the metric can be ignored.
	
	The two-velocity's normalization condition $\dot{x}_\nu \dot{x}^\nu = -1$, for massive particles, implies that
	$1 = f \dot{t}^2 - \dot{r}^2/f$. Substituting the above expressions leaves the two solutions $\dot{t} =  \sqrt{\pi_r^2 + 1/f}$. Using this, the equations of motion of $\boldsymbol{\rho} = (r,\pi_r)$ are $\p_{t} \boldsymbol{\rho} = \dot{\boldsymbol{\rho}}/\dot{t} =  \mathbf{F}(\boldsymbol{\rho})$,
	with
	\begin{align*}
		F_1(\boldsymbol{\rho}) &= \frac{\pi_r f}{\sqrt{\pi_r^2 + 1/f}},\\
		F_2(\boldsymbol{\rho}) &= - \frac{ \pi_r^2/f'}{2  \sqrt{\pi_r^2 + 1/f}}  - f' \frac{\sqrt{\pi_r^2 + 1/f}}{2}  -Q_{\text{e}} A_t'.
	\end{align*}
	We next linearize these equations around the fixed point $\boldsymbol{\rho}_0$, $\mathbf{F}(\boldsymbol{\rho}_0)=0$, to first order
	$\mathbf{F}(\boldsymbol{\rho}) = K(\boldsymbol{\rho_0}) (\boldsymbol{\rho} - \boldsymbol{\rho_0})$,
	where
	\begin{equation}
		K(\boldsymbol{\rho_0}) =\mat{\p_r F_1 & \p_{\pi_r} F_1\\ \p_r F_2 & \p_{\pi_r} F_2}\bigg\vert_{\boldsymbol{\rho} = \boldsymbol{\rho_0}}
	\end{equation}
	is the  Jacobian matrix. Slight perturbations away from a fixed point the dynamics is described by $\boldsymbol{\rho} = e^{t K(\boldsymbol{\rho}_0)}\boldsymbol{\rho}_0$. In terms of the phase space $(r,\pi_r)$, we have a fixed point at $\pi_r=0$ and for massive particles the additional condition that
	\begin{equation}
		Q_{\text{e}} = - \frac{f'(r_i)}{2 f(r_i)^{1/2} A_t'(r_i)}. \label{fixedPointCond}
	\end{equation}
	From here we can either find the corresponding initial $r_i$ given a charge $Q_{\text{e}} $, or we can just consider any $r_i$, but set the charge accordingly. The results are equivalent, but the analysis is simpler for the latter. For massive particles, the matrix $K$ is off-diagonal $K_{11} = K_{22}=0$, with $K_{12}= f^{3/2}$ and
	\begin{equation}
		K_{21}=  f^{-3/2}\left[(f'/2)^2 - Q_{\text{e}} A_t^{(2)} f^{3/2} - f f^{(2)}/2\right].
	\end{equation}
	It has eigenvalues $\lambda_{\pm} = \pm \sqrt{\det{K} }$, where the largest eigenvalue is the Lyapunov exponent $\lambda_+$. To get a measure of how much MSS bound is saturated, we focus on $v_{\text{dJT}} \equiv \lambda_+/\kappa$, explicitly given by
	\begin{align}
		v_{\text{dJT}}&= \frac{f^\prime(r_i)}{f^\prime(r_{\text{H}})}\sqrt{
			{1+\frac{2f(r_i)}{f^\prime(r_i)}}\left[\frac{A_t^{(2)}(r_i)}{A_t^\prime(r_i)}-\frac{f^{(2)}(r_i)}{f^\prime(r_i)}\right]}, \label{LE}
	\end{align}
	which is $1$ if the system is maximally chaotic, in the sense of saturating the MSS bound.  
	
	Using the near horizon emblackening factor, we find \eqref{NHf}
	\begin{equation}
		\frac{2f(r_{\text{H}}+\delta)}{f^\prime(r_{\text{H}}+\delta)} \sim \delta \frac{2 C_\mu  + \delta}{C_\mu+ \delta }, \quad
		\frac{f^{(2)}(r_{\text{H}}+\delta)}{f^\prime(r_{\text{H}}+\delta)} \sim \frac{1}{C_\mu + \delta}. \label{termsInV}
	\end{equation}
Let us further assume that
\begin{equation}
	\frac{A_t^{(2)}(r_{\text{H}}+\delta)}{A_t^\prime(r_{\text{H}}+\delta)} = \frac{1}{\Phi_{\text{th}}'(r_{\text{H}})/\Phi_{\text{th}}^{(2)}(r_{\text{H}}) + \delta} \label{assum}
\end{equation}
is of order $\Oo(\delta^0)$. If we now take the limit as $\delta\to 0$, without specifying any dependent on $q,T_{\text{H}}$ we get one of two results. For $T \neq 0$, we have a non-extremal black-hole and 
	$v_{\text{dJT}}(\delta)=\sqrt{1+\Oo(\delta)} \to 1$.
In other words, at finite $\beta$, we obtain a Lyapunov exponent saturating the MSS bound. This is in both phases, which agrees with the Lyapunov exponents of the gaseous and liquid phases in the rescaled regime $T = q^{-1} \tilde{T}$, $\mu = q^{-3/2} \tilde{\mu}$  of the cSYK model \cite{Louw2023}.

 \subsection{Near-extremal case}
We now wish to compare to the results in the second rescaled regime $T = q^{-2} \bar{T}$, $\mu = q^{-2} \bar{\mu}$.
 As $T_{\text{H}} \to 0$, so does the specific heat, meaning that 
	\begin{equation}
		\frac{2f(r_{\text{H}}+\delta)}{f^\prime(r_{\text{H}}+\delta)} \to \delta, \quad
		\frac{f^{(2)}(r_{\text{H}}+\delta)}{f^\prime(r_{\text{H}}+\delta)} \to \frac{1}{\delta}.
	\end{equation} 
	With this \eqref{LE} reduces to   
	\begin{equation}
		v_{\text{dJT}}(\delta)=\sqrt{1+\delta[\Oo(\delta^0)-\delta^{-1}]} \to 0.
	\end{equation}
	corresponding to an extremal black-hole with emblackening factor $f(r)= f^{(2)}(r_{\text{H}}) \delta^2/2$. An exception to the above occurs if the electrical potential contribution leads to a perfect cancelation such that $v$ remains equal to $1$. We have assumed that \eqref{assum} remains of order $\delta^0$. To assess the validity of this assumption we calculate \eqref{assum} for both possible dictionaries \ref{Dictionary2}.a and \ref{Dictionary2}.b. We write this as deviations from the specific heat  
\begin{equation}
    C^{(a/b)} \equiv \frac{\Phi_{\text{th}}^{(1)}(r_H)}{\Phi_{\text{th}}^{(2)}(r_H)} = C_{\mu} - \delta_{\pm}^{(a/b)}\label{PhiPhi}
\end{equation}
Here we use the $a/b$ to denote the cases given the two dictionaries in Table. \ref{Dictionary2}. In this notation, a perfect cancelation will occur if $\delta_{+}^{(a/b)}$ goes to zero. Here $\Phi_{\text{th}} \propto \Qq-1/2$ for dictionary \ref{Dictionary2}.a and $\Phi_{\text{th}} \propto E = H + (\Qq-1/2)\mu$
for dictionary \ref{Dictionary2}.b. With this, we have 
$$\Phi_{\text{th}}^{(1)}(r_H) = \begin{cases}
    \Qq'(r_H) & (a)\\
    T + \mu \Qq'(r_H) & (b)
\end{cases}$$
where we have used the enthalpy relation $\p_{\Ss} H = T$. Further, recalling that $C_\mu = T \p_T \Ss$, we have the second derivatives
$$\Phi_{\text{th}}^{(2)}(r_H) = \begin{cases}
    \Qq^{(2)}(r_H) & (a)\\
    T/C_\mu + \mu \Qq^{(2)}(r_H) & (b),
\end{cases}$$
where $\Qq^{(2)}(\Ss) = -\Ss^{(2)}(\Qq)/\Ss'(\Qq)^3$. For the non-interaction system we find $\Ss_0' = - \bar{\beta} \bar{\mu}$ and $\Ss^{(2)}_0 = - (\bar{\beta} \bar{\mu})^2 C_{\mu}^{(0)}$. As such, without any interactions, one finds that \eqref{PhiPhi} is exactly equal to $C_{\mu}^{(0)}$, in other words, the same term as in \eqref{termsInV}. However, there are still contributions stemming from the interactions. Now for the lower boundary $\Qq \to 1/2$, where $C_\mu^{(0)} \sim 2(\bar{\beta} \bar{\mu})^2 e^{-\bar{\beta}\bar{\mu}}$,
$$\delta_{-}^{(a)} \sim \frac{(\pi v/2)^2}{2}(\bar{\beta}\bar{\mu})^2,\quad \delta_{-}^{(b)} \sim C_\mu^{(0)}/2 $$
and for the upper boundary $\Qq \to 1/q$, we find
\begin{equation}
    \delta_{+}^{(a/b)} \sim 2\frac{v-1}{q}+ \frac{2(2+\pi^2)/q^2}{\pi^4 -4 \pi^2 -2} \label{delta+}
\end{equation}

From the above, a perfect cancelation in the larger black hole if we first take the $q\to\infty$ limit in \eqref{delta+}. This then implies that the large black hole is still maximally chaotic in the sense that $v_{dJT} \to 1$. This result would then match with the gaseous Majorana-like ($\Qq=0$) SYK phase at low temperature.  

The same can happen in the smaller black hole depending on how the limit is taken. The smaller black hole seems rather badly behaved in terms of the emblackening factor. Especially when considering the black hole charge to be the conjugate driving the phase transition, one should also consider a possible free AdS phase. In other words, one should perform a similar analysis to that of Hawking and Page \cite{Hawking1982Jan}, examining the free energy of the pure AdS solutions to determine when and how this crossover occurs. This would modify the interpretation of the low-temperature regime.

Given the above analysis, one should also note its possible limitation. This lies in the fact that for the extremal black hole the charge of the test particle \eqref{fixedPointCond} tends to diverge at the fixed point. For such a diverging charge, it is unlikely that one can ignore the back-reaction from the charged particle \cite{Poisson2004Dec}.

\section{Conclusion}

Previous analogies between  RN-AdS black holes with spherical event horizons and the van der Waals liquid past  due to their similar phase structure \cite{Kubiznak2012Jul}. However, on the dual field theory side, there is a lack of equivalent holographic descriptions in the literature. In this work, we provided such a holographic description between the $(0+1)$-dimensional cSYK model and $(1+1)$-dimensional JT gravity with a particular deformation. In particular, we have provided a deformed JT gravitational model with a matching partition function to the $q/2$-body interacting cSYK model for large $q$. Moreover, together with matching equations of states, we have an exact equivalence in the thermodynamics. We achieved this by introducing a deformed JT gravity model characterized by a dilaton potential $\Uu(\varphi)$ and dilaton-to-Maxwell field coupling $\Ww(\varphi)$, and deriving the black hole metric in terms of the physical quantities of the cSYK model. 

One of the original reasons for believing that the SYK model should have a holographic dual was its maximal Lyapunov exponent, which is also found in gravitation theories \cite{Kitaev2015}. As such, we went beyond the thermodynamic description and also considered the chaotic nature of the black hole. Close to the second-order phase transition, both liquid and gaseous phases of the cSYK model are maximally chaotic. We estimated the Lyapunov exponent on the gravitational side via linear stability analysis. This indicated the standard maximal Lyapunov exponents associated with both large and small black hole phases. As such, in the first rescaled regime of the phase diagram, we not only found the same thermodynamics but also the same Lyapunov exponents.

It is known that the Lyapunov exponents of black holes in the extremal limit tend to zero. This is a side effect of the bound $2\pi T$ tending to zero since the extremal limit corresponds to the zero temperature limit. As such we focused on the ratio $v = \lambda_L/(2\pi T)$. For the cSYK model, the liquid phase would remain maximally chaotic $v=1$, while the gaseous phase becomes regular $v\to 0$. Depending on the choice of dictionary, and how the limits are taken, one can get different results for the small and large black holes. This highlights the need for a more in-depth analysis taking the black hole back action into account. Open questions also remain in terms of the appropriate UV and IR cutoffs to prevent unphysical behavior in the black hole. As an example, for ordinary $(1+3)$-dimensional black hole chemistry the smaller black hole's radius does not shrink to zero \cite{Kubiznak2012Jul}. When setting the charge to zero, the black hole no longer exists at $T=0$. A natural question is whether some interpretation changes could yield similar results.

The provided dictionaries directly overlap with the analogies between the charged SYK model and charged black holes provided in \cite{Louw2023}. As such this paper directly serves as an answer to said paper, by both showing that the analogies can be used as dictionaries.
In conclusion, our results encourage the use of holography away from the low-temperature regime, i.e., beyond the near-extremal regime.

\section*{Acknowledgements}

We would like to thank Wenhe Cai, Yicheng Rui, and Rishabh Jha for their helpful discussions. This work is partly supported by NSFC, China (Grant No. 12275166 and No. 11875184) and partly by the Deutsche Forschungsgemeinschaft (DFG, German Research Foundation) - SFB 1073.
	
	\bibliography{refs.bib}
	
	\appendix

	\section{Extremizing the action \label{secEMDaction}}
	In this section, we find and solve the Euler-Lagrange equations associated with the deformed JT gravity action
	$$I[\varphi,A] = \frac{1}{G_{\text{N}}}\nint[M]{^2 x} \sqrt{-g} \Ll(\varphi, A) + I_{\text{bdy}},$$
	with boundary action contribution $I_{\text{bdy}}$, described in sec. \ref{AppFreeEnergy}, which cancels any divergences. The bulk Lagrangian \eqref{BulkLag} is given by 
	\begin{equation}
		\Ll(\varphi, A) = \frac{\varphi}{4\pi} \Rr_2 +P \Uu(\varphi) - \frac{\Ww(\varphi)}{4}  F(A)^2,\label{LlBulk}
	\end{equation}
	with dimension $\ell^{-2}$ and a dimensionless dilaton field $\varphi$.
	We assume the black hole solution, in the Schwarzschild gauge, takes the form $ds^2 = - f(r) dt^2 + dr^2/f(r)$ in Lorentzian signature. For such a metric, in $(1+1)$-dimensions, the Ricci scalar takes the form
	\begin{equation}
		\Rr_2(r)= -f^{(2)}(r) \label{Ricci2}.
	\end{equation}
	
	Varying with $A$ yields $\nabla_{\mu} \Ww F^{\mu\nu}$. Due to the symmetry of the Christoffel symbols and the asymmetry of $F_{\mu\nu} = 2\p_{[\mu} A_{\nu]}$, this reduces to $\p_\mu \Ww F^{01}=0$ which is solved by
	\begin{equation}
		F^{01} =  Q_{\text{B}}/\Ww, \quad F^2 = -2 Q_{\text{B}}^2/\Ww^2. \label{F^{01}}
	\end{equation} 
	
	Varying with $\varphi$ yields $P\Uu'-\Ww' F^2/4 = -\Rr_2/(4\pi)$, which together with the on-shell relation \eqref{F^{01}} and \eqref{Ricci2} becomes
	\begin{equation}
		f^{(2)}(r) = 2\pi \p_{\varphi}[2 P \Uu(\varphi) - Q_{\text{B}}^2/\Ww(\varphi)]. \label{f2}
	\end{equation}
	
	Noting that the Einstein tensor is zero in $2$-dimensions, varying with $g$ leaves only \cite{Cai2020Aug}
	\begin{align}
		\nabla_\mu \nabla_\nu \frac{\varphi}{\pi}  &=  g_{\mu\nu} \left[\nabla^2 \frac{\varphi}{\pi} - 2 P \Uu + \Ww \frac{F^2}{2} \right] - 2\Ww F_{\mu\rho} F^{\rho}_{\nu} \label{varg}
	\end{align}
	Here we have used the identity \cite[20.22]{BibEntry2020Dec}
	$$\delta \Rr_2 = [\Rr_2]_{\mu\nu} \delta g^{\mu\nu} - (\nabla_\mu\nabla_{\nu} - g_{\mu\nu} \square) \delta g^{\mu\nu}$$
	and used integration by parts.	Using \eqref{f2}, the dilaton equation \eqref{varg} reduces to the coupled set of differential equations
	\begin{align}
		f'(r) \p_r^2 \varphi =- 2 \p_r [f(r)\p_r^2 \varphi], \quad
		\p_\tau^2\varphi =  f(r)^2 \p_r^2 \varphi \label{dil2}
	\end{align} 
	where we have performed a Wick rotation to imaginary time $t \to \i \tau$. Together, these equations 
	reduce to $\p_r^2 \varphi = 2c f(r)^{-3/2}$ and $\p_\tau^2\varphi = 2 c \sqrt{f(r)}$, for some constant $c$. The latter equation is solved by
	\begin{equation}
		\varphi(\tau,r) = \tau (b  + c \tau) \sqrt{f(r)} + R(r), \label{phiSol}
	\end{equation}
	for some $\tau$ independent function $R(r)$ and constants $b,c$. Inserting \eqref{phiSol} into the former equation $f(r)^{3/2} \p_r^2 \varphi = 2 c$, 
	\begin{equation}
		2 c = f(r)^{3/2} R^{(2)}(r) - \tau (b+ c \tau)\frac{f'(r)^2 - 2 f(r) f^{(2)}(r)}{4}. \label{JTeq}
	\end{equation}
	Note that \emph{only} the last term in \eqref{JTeq} has time dependence. Since this equation should hold for all $\tau$, these $\tau$ dependent parts must cancel, i.e., either $b=c=0$ or $	f'(r)^2 = 2 f(r) f^{(2)}(r)$. The latter equation is solved by $f_{\text{ext}}(r) = z^{-2}$, where $z = m/(r-r_{\text{H}})$, where $m$ is some constant and $r_{\text{H}}$ is the event horizon radius. In the JT case where $\Ww = 0$ (or the charge $Q_{\text{B}}=0$) and $\Uu = \varphi$, we would find that 
	\begin{equation}
		f_{\text{JT}}(r) = 2\pi P(r-r_+)(r-r_-). \label{JTf}
	\end{equation}
	Hence, we note that the solution $f_{\text{ext}}(r)$ is that of an extremal (zero temperature) black hole $r_\pm \to r_{\text{H}}$. However, as $r\to\infty$, we also find that $f_{JT}(r) \to f_{\text{ext}}(r)$. This leaves the most general extremal solution to the dilaton \eqref{phiSol}
	\begin{equation}
		\varphi_{\text{extr}}(\tau,r) = \frac{a + b \tau + c [\tau^2 + m^2 z^2]}{z} + d.
	\end{equation}
	
	
	In general, however, for non-zero temperatures, the solution must therefore have $c = b =0$, i.e., $\varphi(r) = R(r)$, with \eqref{dil2} indicating that $\varphi^{(2)}(r) = 0$, which is solved by
	\begin{equation}
		\varphi(r) = \gamma r + \varphi_0 \label{dilSol}
	\end{equation} 
	for some coupling strength $\gamma$. For such a time-independent solution,	\eqref{F^{01}}
	\begin{equation}
		F^{01} = Q_{\text{B}}/\Ww(\varphi) = \p_t A_{r}-\p_r A_{t} \label{rel1}
	\end{equation}
	is also time-independent. As such, we may choose the gauge $\p_t A_{r} = 0$, meaning that $A_t$ is the antiderivative 
	\begin{equation}
		A_t(r) = \frac{Q_{\text{B}}}{\gamma}\nint{\varphi} \frac{1}{\Ww(\varphi)}. \label{antiDerA}
	\end{equation}

	\subsection{Emblackening factor solutions in the non-extremal case}
	From \eqref{dilSol}, $\gamma dr =  d\varphi$, and we may integrate \eqref{f2} over $r$ to yield
	\begin{equation}
		\gamma f'(r)/(2\pi) = 2 P \Uu(\varphi) - Q_{\text{B}}^2/\Ww(\varphi) + 2  T_0. \label{EOS1}
	\end{equation}
	Here we have allowed for the possibility of some integration constant $4\pi  T_0$ would amount to a shift in temperature $T_0$ and adding a linear term $T_0 \varphi_0$ to the enthalpy. We will later see that this term has no effect on the physics. Integrating once more yields
	\begin{equation}
		\gamma f(r)/(2\pi) = 2 P V(r) - Q_{\text{B}} A_t(r)+ 2 T_0 r - 2 M, \label{f}
	\end{equation}
	where $M$ is some to-be-determined/interpreted integration constant, and we have defined the anti-derivatives
	\begin{equation}
		V(r) = \gamma^{-1} \nint{\varphi} \Uu(\varphi), \quad \Phi(r) = -A_t(r),
	\end{equation}
	where $\Phi(r) = -A_t(r)$ is the electrical potential at $r$, w.r.t. the horizon.
	Further, at the horizon $r_{\text{H}}$, $f(r_{\text{H}}) = 0$, we find
	\begin{equation}
		M = P V_{\text{th}} + Q_{\text{B}} \Phi_{\text{th}}/2 + T_0 r_{\text{H}}, \label{ADMmass}
	\end{equation}
	where $V_{\text{th}} \equiv V(r_{\text{H}})$ is the thermodynamic volume and $\Phi_{\text{th}} \equiv \Phi(r_{\text{H}})$. Setting $\gamma = 1$, amounts to measuring $r$ in units of $\gamma$.
	
	\subsection{Free energy \label{AppFreeEnergy}}
	While working in Euclidean signature $\tau_{\text{E}} = \imath t$, the periodicity $\beta_H = 4\pi/f'(r_{\text{H}})$ in the metric is required to avoid a conical singularity.  We associate the free energy with the on-shell action $ \Ff/G_{\text{N}} =  I_E^*/\beta_H + I_{\text{bdy}}^*/\beta_H$.  Substituting the on-shell solutions into the bulk Lagrangian density \eqref{LlBulk} yields
	\begin{equation}
		\Ll = -\frac{r+\varphi_0}{4\pi}f^{(2)}(r) +P V'(r) - Q_{\text{B}} \Phi'(r)/2.\label{LlBulk2}
	\end{equation}
which may be rewritten as
	\begin{equation}
		\Ll = -\frac{ [(r+\varphi_0) f^{(1)}(r)-f(r)]' }{4\pi}+ \p_r [P V - Q_{\text{B}}  A_t/2]
	\end{equation}
	Further, the Euclidean action takes the form
	\begin{equation}
		G_{\text{N}} I_E^* = -\nint[0][\beta_{\text{H}}]{\tau} \nint[r_{\text{H}}][r_{\text{max}}]{r} \Ll
	\end{equation}
	which, together with \eqref{GHY} and \eqref{ct}, leaves us with the on-shell action
	\begin{equation}
		\Ff = -\frac{ [r_{\text{H}}+\varphi_0] f'(r_{\text{H}})- f(r_{\text{H}}) }{4\pi}+ P V_{\text{th}}- Q_{\text{B}}  \Phi_{\text{th}}/2 + C \label{Ff}
	\end{equation}
	We leave the proof that the appropriate boundary terms allow us to set $C=0$, for App. \ref{sec:Counter}. Noting $f'(r_{\text{H}}) = 4\pi T_{\text{H}}$ allows us to identify the conjugate to the temperature $\Ss_{\text{W}} = r_{\text{H}}+ \varphi_0$ with the Wald entropy. Since we expect zero entropy when the black hole evaporates to zero $r_{\text{H}}\to 0$, we set $\varphi_0=0$. Lastly since $f(r_{\text{H}}) = 0$, we are then left with the free energy
	\begin{equation}
		\Ff = M - Q_{\text{B}} \Phi_{\text{th}} - T_{\text{H}} \Ss_{\text{W}}.\label{UJT}
	\end{equation}
	Having identified the entropy and temperature, one may show that $M$, defined in \eqref{ADMmass}, is the mass \cite{Cai2020Aug,Gegenberg1995}. From this, one may obtain the Hawking temperature $T_{\text{H}} \equiv 1/\beta_H$ as
	
	\begin{equation}
		\left(\frac{\p M}{\p \Ss_{\text{W}}}\right)_{Q_{\text{B}},P}= T_{\text{H}} = P V_{\text{th}}'(\Ss_{\text{W}}) + Q_{\text{B}} 
		\Phi_{\text{th}}'(\Ss_{\text{W}})/2 + T_0, \label{EOST1}
	\end{equation}
	Together \eqref{UJT} and \eqref{EOST1} define the thermodynamics of the system. We note that $T_0$ shifts the definition of temperature. Due to our freedom in choosing the potentials defining the temperature relation, we may set $T_0=0$ without loss of generality.
	
	\subsection{The Gibbs free energy and the uncharged dual} 
	In black hole chemistry, the pressure and thermodynamic volume are conjugate to another \cite{Dolan2016May}. This leads to the identification of the ADM mass with the enthalpy, indicating that $Q_{\text{B}} \Phi_{\text{th}}/2$ is the interaction energy. Further, the thermodynamic potential which selects out the favorable state is the Gibbs free energy \cite{Kubiznak2017Feb} 	
    \begin{equation}
		G = M - T_{\text{H}} \Ss_{\text{W}}, \label{Gibbs}
	\end{equation}
    with differential $dG = -\Ss_{\text{W}} dT_{\text{H}} + V_{\text{th}} dP + \Phi_{\text{th}} d Q_{\text{B}}$.

    Let us now consider the uncharged case by setting the term $\Ww(\varphi) F(A)^2$ equal to zero in \eqref{BulkLag}. Following this, we make the replacement $	P \Uu \to P \Uu - Q_{\text{B}}^2/(2\Ww)$. One may note that this leaves all the equations of motion the same as the charged case \cite{Grumiller2007Apr,Cai2020Aug}. Such a replacement is also equivalent to replacing the varying Maxwell field with its on-shell part in the action.  This replacement does, however, yield a single difference---a sign flip in the above on-shell action \eqref{UJT} in the charged term
	\begin{equation}
		\frac{G_{\text{N}} I^*_{\text{uncharged dual}}}{\beta_{\text{H}}} = P V_{\text{th}} + Q_{\text{B}} \Phi_{\text{th}}/2 - T_{\text{H}} \Ss_{\text{W}}.\label{IGibbs}
	\end{equation}
	Recalling the expression for the ADM mass \eqref{ADMmass} we note that \eqref{IGibbs} is the Gibbs free energy $G$. This is somewhat reminiscent of the relations between the canonical ensembles. In the first case, we originally allowed the electromagnetic field to vary, leading to the $\Ff$. Instead, replacing the Maxwell field by its on-shell part, we no longer treat it as its own independent parameter. This then yields the Gibbs free energy.
 
\section{Low energy effective action and counter  terms \label{sec:Counter}}

 \subsection{Gravitational boundary term}
  Here we show how the appropriate boundary term counters perfectly the divergences to yield a constant part in \eqref{Ff}
	\begin{equation}
		C = \frac{\varphi_{\text{max}} f'(r_{\text{max}})- f(r_{\text{max}}) }{4\pi}- [P V - Q_{\text{B}} \Phi/2]\big\vert_{r=r_{\text{max}}}+\frac{I_{\text{bdy}}}{\beta_{\text{H}}} \label{constPart}
	\end{equation}
	which is perfectly canceled by the boundary action $I_{\text{bdy}} = I_{\text{ct}} +I_{\text{GHY}}$.
	This is composed of two parts, counter and the Gibbons-Hawking-York (GHY) terms given by
	\begin{equation}
		I_{\text{ct}} \equiv \frac{1}{G_{\text{N}}}\nint[\p M]{\tau} \sqrt{h} \Ll_{\text{ct}}, \quad I_{\text{GHY}} \equiv -\frac{1}{G_{\text{N}}}\nint[\p M]{\tau} \sqrt{h} \varphi K, \label{bdyTerms}
	\end{equation}
	respectively. The induced metric and extrinsic curvature of the boundary $r=r_{\text{max}}$ entering the above are given by $h = f(r_{\text{max}})$ \cite{Cao2021Mar} and $K = -\p_{r_{\text{max}}} \sqrt{h}$ respectively, yielding \eqref{bdyTerms}
	\begin{align}
		\frac{G_{\text{N}} I_{GHY}^* }{\beta_{\text{H}}}
		&= -\frac{\varphi(r_{\text{max}}) f'(r_{\text{max}})}{4\pi}. \label{GHY}
	\end{align}
	The on-shell counter Lagrangian, on the boundary, we require is 
	$$\Ll_{\text{ct}} = \frac{ \sqrt{f(r_{\text{max}})}}{4\pi} + \frac{P V - Q_{\text{B}} \Phi/2}{\sqrt{f(r_{\text{max}})}},$$
	which yields the on-shell contribution 
	\begin{align}
		\frac{G_{\text{N}} I_{\text{ct}}^*}{\beta_{\text{H}}} 
		&= \frac{ f(r_{\text{max}})}{4\pi} + [P V - Q_{\text{B}} \Phi/2]\big\vert_{r=r_{\text{max}}}.\label{ct}
	\end{align}
	Substituting \eqref{GHY} and \eqref{ct} into \eqref{constPart}, we see how the divergences cancel to zero $C = 0$.

\section{cSYK entropy form \label{twoStateEntropy}}
	
	In this section, we show that the complex SYK model has an entropy of the form
	\begin{equation}
		\boxed{\Ss(\Qq) =	\Ss_{2}(\Qq) -(\pi v/q)^2/2.} \label{Scor}
	\end{equation}
	We do this by using the Maxwell relation
	\begin{equation}
		-\left(\frac{\p \Ss}{\p \Qq} \right)_{T,J} = \left(\frac{\p \mu}{\p T}\right)_{\Qq,J},
	\end{equation}
	together with the EOS
        \begin{equation}
		T  = \frac{\mu - 4\Qq \epsilon/q}{2\tanh^{-1}(2\Qq)}, \label{TSYKEOSA}
	\end{equation}
        rewritten in the form
	\begin{equation}
		\mu= 2 T \tanh^{-1}(2\Qq)+ 4 \Qq \epsilon/q, \label{EOSAp}
	\end{equation}
	where $\epsilon \equiv \Jj(\Qq) \sin(\pi v/2)$ and the non-interacting part corresponds to $\Ss_2'(\Qq) = -2 \tanh^{-1}(2\Qq)$. As such, we are left to show that the corrections $\Ss_{I} = \Ss-\Ss_2$ satisfy
	\begin{equation}
			-q\left(\frac{\p \Ss_{I}}{\p \Qq} \right)_{T,J} = \left(\frac{\p \Qq \epsilon}{\p T} \right)_{\Qq,J}
	\end{equation}
or explicitly, for \eqref{Scor} to hold, we must show that 
	\begin{equation}
	(\p_\Qq v)_{T,J} \pi^2 v/q = 2\pi \Qq \Jj(\Qq) \cos(\pi v/2) (\p_T v)_{\Qq,J}.
	\end{equation}
From the closure relation \eqref{clos}
\begin{equation}
	\beta\Jj(\Qq) = \pi v \sec(\pi v/2) \label{clos2}
\end{equation}
 this relation reduces to $	(\p_\Qq v)_{T,J}/q = 2 \Qq T (\p_T v)_{\Qq,J}$. To find the derivatives of $v$, we differentiate both sides of the closure relation with respect to some variable $x$
$$\p_x \ln [\beta\Jj(\Qq)] = \p_x \ln[\pi v \sec(\pi v/2)] = a(v) \p_x \ln v,$$
where $1/b(v) \xrightarrow{v\to 0} v$ and $1/b(v) \xrightarrow{v\to 1} 1-v$ is explicitly given by $b(v) = a(v)/v$ with
$$a(v) \equiv 1 + \pi v \tan(\pi v/2)/2 = 1 + \beta \epsilon/2.$$ 
So we have $(\p_\Qq v)_{T,J} = b(v)^{-1}\p_\Qq \ln \Jj(\Qq)$ and $(\p_T v)_{\Qq,J} = -b(v)^{-1}\p_T \ln T$. As such, we only have left to show that $	\p_\Qq \ln[ \Jj(\Qq)]/q = -2 \Qq$, which follows from the definition of effective interaction, for small charge densities, $\ln \Jj(\Qq) \sim -q \Qq^2$. For larger charge densities, the SYK contribution is exponentially suppressed in $q$ and the theory is only described by the non-interacting part. As such, we have shown that \eqref{Scor} holds.
	
	One may show that this entropy remains correct even for zero charge density. This may be verified using the differential \eqref{dOmega} at $\mu=\Qq=0$ together with the closure relation \eqref{clos2}.

	\subsection{Inverse function of two-state entropy \label{IntSYKrel}} 
	
	Here we find the inverse function $r = \Ss_2^{-1}(x)$ to the two-state entropy function \eqref{entExpl}
	\begin{equation}
		\Ss_{2}(x) \equiv - \frac{1-2 x}{2} \ln\!\left\vert\frac{1-2 x}{2}\right\vert- \frac{1+2 x }{2} \ln\!\left\vert\frac{1+2x}{2}\right\vert.
	\end{equation}
	
	For $x= 1/2-\sigma$, $\Ss_{2}(x)  = \sigma [1-\ln \sigma] + \Oo(\sigma^2)$, where $r = \sigma_{1/2} [1-\ln \sigma_{1/2}]$ is solved by $\sigma_{1/2} =e^{1+\Ww_{-1}(-r/e)} $, which is the product log function \cite{Fedoryuk1989} 
	\begin{equation}
		\Ww_{-1}(-r) = \ln(r)-\ln(-\ln(r)) +\Oo\left(\frac{\ln(-\ln(r))}{\ln(r)}\right) \label{liqXsim}
	\end{equation}
With this, the solution may be written as $\sigma_{1/2}(r) \sim -r/\ln r$. Around $x=0$,	$\Ss_2(x) = \ln2 -2x^2 + \Oo(x^4)$, solved by
	\begin{align}
		x(r) 
		&= \sqrt{[\ln2-r]/2} + \Oo(\ln2-r)^{3/2}.
	\end{align}

		Close to the IR cutoff $x(r\to r_{\text{max}}) = x_\text{min}$, for $x = \Oo(q^{-2})$, we should consider the full entropy function \eqref{Scor} 
		$$\Ss(x) = \ln 2 - \frac{4 \bar{x}^2 + (\pi v)^2}{q^2} + \Oo(q^{-4})$$ 
		Assuming $\Tt(x) = o(q^0)$, we have $v(x) = 1 + o(q^0)$. Here we are using little o notation, where $o(q^0)$, means sub-leading in $q^0$, e.g., $1/\ln q$. In terms of the radii we have the equation
		$r(x) \sim r_{\text{max}} + 2(x_{\text{min}}^2 - x^2)$, which is solved by
		\begin{equation}
			x(r) \sim \sqrt{x_{\text{min}}^2 + (r_{\text{max}}-r)/2  } \label{xcutoff}
		\end{equation}

\subsection{Specific heat \label{AppSpecificHeat}}

The typical Majorana SYK model has thermodynamics matching the cSYK model at half-filling $\Qq=0$. For this case, we note that the entropy is merely given by
$\Ss = \ln 2 -(\pi v/q)^2/2$. The corresponding specific heat $C_\mu = T \p_T\Ss$, is then found by considering how $v$ changes with temperature, as described by the closure relation \eqref{clos2}. Such an analysis eventually reveals $C_\mu \sim 2 (\pi/q)^2 T/J$ as $T \to 0$. Since the equation of state, as written in \eqref{TSYKEOSA}, is no longer valid for $\Qq=0$, a natural question is whether this specific heat can still be obtained from the general analysis. Here we show that this is indeed the case. 

For the full specific heat, we consider 
$$\left(\frac{\p \Ss}{\p T}\right)_{\mu,J} = \left(\frac{\p \Qq}{\p T}\right)_{\mu,J}\Ss_2'(\Qq) - \left(\frac{\p v}{\p T}\right)_{\mu,J} \left(\frac{\pi}{q}\right)^2 v.$$

As before, we will assume, unless stated otherwise, that $\mu,J$ is kept constant. We find relate $\p_T v$ to $\p_T \Qq$, by considering both the EOS and the closure relation \eqref{clos2} yielding
$$\left(\frac{\p v}{\p T}\right)_{\mu,J} = -\frac{\beta + 2 q \Qq \p_T \Qq}{b(v)}$$
and 
$$\left(\frac{\p \Qq}{\p T}\right)_{\mu,J} \sim\frac{1-4\Qq^2}{4} \frac{\beta \Ss_2'(\Qq) - \Qq  \p_T (\pi v)^2/q}{1+(1-2q\Qq^2) \beta\epsilon/q}$$
which together yields
$$\left(\frac{\p \Qq}{\p T}\right)_{\mu,J} = \frac{1-4\Qq^2}{4 T}\frac{ \Ss_2'(\Qq) +2 \pi^2 \Qq [v/b(v)]/q}{   1+\beta\epsilon/q- \Qq^2 [2\beta\epsilon+\pi^2 v/b(v)] }.$$
The general expression reads
\begin{equation}
    q^2 C_{\mu} = T\left(\frac{\p \Qq}{\p T}\right)_{\mu,J}\left[q^2\Ss_2'(\Qq) + 2 q \Qq \frac{\pi^2 v }{b(v)} \right]+\frac{\pi^2 v}{b(v)}. \label{genC}
\end{equation}
As $\Qq\to 0$, only the final expression $\pi^2 v/b(v) \to 2 \pi^2 T/J$ remains
where $v = 1-2 T/J + \Oo^2(T/J)$, hence specific heat $C_\mu \sim 2 (\pi/q)^2 T/J$. This corresponds to a large negative curvature $\Rr_2(\ln 2) \sim -2 J q^2/\pi$.

\section{IR cutoff details \label{detailsIRCutoffs}}

We choose this cutoff such that our theory will satisfy two conditions:

\emph{(I)}	Given the cutoff we have access to the full coexistence line of the SYK model. 

\emph{(II)}	The scalar curvature $\Rr_2(r_{\text{max}})$ remains finite for any finite value of $q$. This condition would, for instance, be violated given an emblackening factor $f(r) \propto \sqrt{r_{\text{max}}-r}$, which has both a diverging temperature function (related to $f'(r)$) and scalar curvature (related to $f^{(2)}(r)$).

	With these conditions in mind, let us consider the radius as a function of $x_{\text{min}}$ when $x_{\text{min}} = \bar{x}_{\text{min}}/q$. This can be done since the entropy function relates these two $r = \Ss(x)$. Expanding the entropy function around $\bar{x}_{\text{min}} \le \Oo(q^{0})$, we find 
\begin{equation}
	r_{\text{max}} = \ln 2 - \frac{4 \bar{x}_{\text{min}}^2 +(\pi v(x_{\text{min}}))^2+ \Oo(q^{-2})}{2q^2} .
\end{equation}
Note that this maximal event horizon radius tends to the maximal entropy of a two-state system $r_{\text{max}} = \Ss_{\text{max}} \xrightarrow{q\to\infty} \ln2$, which is also the von Neumann entropy of maximally entangled Bell states. 

\subsection{Condition (I): Physics along the coexistence line}
For $\bar{x}_{\text{min}}/q$, we will find a temperature function  \eqref{f'(r)} of the order $\Tt(x_{\text{min}}) = o(q^0)$, for $\mu \le \Oo(q^{-3/2})$, which includes the entire coexistence line. Let us see this explicitly. In the second order PT regime $\mu = \tilde{\mu} q^{-3/2}$, yielding
\begin{equation}
	\Tt(x_{\text{min}}) \sim  \frac{q^{-1/2}\tilde{\mu}}{4 \bar{x}_{\text{min}}}- J e^{-\bar{x}_{\text{min}}^2/q}/q \label{f'(r)2}
\end{equation}
which is of the order $\Oo(1/q)$ for the chaotic-to-nonchaotic PT regime, $\mu = \bar{\mu}/q^2$.

For $\Tt(x_{\text{min}}) = o(q^0)$ the closure relation \eqref{clos} yields $v(x_{\text{min}}) = 1 + o(q^0)$. This also shows how, on the quantum side, the maximally large black hole ($r_{\text{H}} = r_{\text{max}}$) corresponds to a maximally chaotic SYK model. This is seen in the Lyapunov exponent $\lambda_L = 2\pi T v$ saturating the MSS bound $2\pi T$ for $v(x(r_{\text{H}})) \to 1$. As such we are left with
\begin{equation}
	r_{\text{max}} = \ln 2 - \frac{4 \bar{x}_{\text{min}}^2 +\pi^2 +o(q^0)}{2q^2}. \label{rmax}
\end{equation}

The second condition can only be violated for certain choices of $x_{\text{min}}$. As example would be to demand that $x_{\text{min}}$ be an exact root of $\Tt$, i.e., when $\mu = 4 x \epsilon(x)/q$. Due to the closure relation \eqref{clos}, we know that this root $(\Tt(x) = 0)$ implies that $v(x) = 1$, i.e., $\mu = 4 x \Jj(x)/q$. For small $x = \tilde{x}/\sqrt{q}$, this yields the equation $\tilde{\mu}/(4J) = \tilde{x} e^{-\tilde{x}^2}$. This is solved by the product log function $\Ww_0$
$$\tilde{x} = \sqrt{\frac{-\Ww_{0}(-\tilde{\mu}^2/(8J^2))}{2}}\xrightarrow{\tilde{\mu}\to 0} \tilde{\mu}/(4 J)$$
which is real for $\tilde{\mu}/J \le (8/e)^{1/2} < 2$. Limiting ourselves to only such values of the chemical potential unfortunately would limit our scope to only the line of first-order phase transitions. This is because the second order phase transition occurs at $\tilde{\mu}_{\text{c}}/J = (6/e)^{3/2} >3$.

There does, however, exist one case where $\Tt(x_{\text{min}})$ should be close to zero. This is when the $r_{\text{H}} = r_{\text{max}}$, i.e., for a near-extremal maximally large black hole. This phase is stable when $\mu_0 = \bar{\mu}_0/q^2$. Here $\bar{\mu}_0 = 4J$ is the endpoint of the coexistence line plotted in Fig. \ref{HPSYKphasediag}. This phase should have a temperature of order $q^{-2}$, $T = \bar{T}q^{-2}$. Substituting this into \eqref{f'(r)2}, for a maximally large black hole, we find 
\begin{equation}
	\Tt(x_{\text{min}}) \sim  
	J\frac{1 - \bar{x}_{\text{min}}(1-\bar{x}_{\text{min}}^2/q)}{\bar{x}_{\text{min}} q}
	\label{f'(r)3}
\end{equation}	
which is of order $q^{-2}$ when
\begin{equation}
	\bar{x}_{\text{min}} =1 + \frac{1 -\bar{y}/J}{q}, \label{xmin}
\end{equation}
for some ``freely'' chosen $\bar{y}$. Substituting this result back into \eqref{f'(r)3}, we find that $\Tt(x_{\text{min}}) = \bar{y}/q^2$. This allows us to consider charge densities up to the minimum charge density which occurs along the coexistence line $x_{\text{min}} = 1/q$ \cite{Louw2023}. Substituting \eqref{xmin} into \eqref{rmax} we are left with 
\begin{equation}
	r_{\text{max}} = \ln 2 - \frac{4 +\pi^2+ \Oo(q^{-2})}{2q^2} .
\end{equation}
which is independent of $\mu/J$. We also have 
\begin{equation}
	\Tt(x_{\text{min}}=1) \sim  q^{-1/2}\tilde{\mu}/4- J/q +\Oo(q^{-3/2})
\end{equation}  
Note that while we have written this equation for the scaling $\mu = \Oo(q^{-3/2})$, it remains valid for $\mu = \Oo(q^{-2})$. In that case, it may be written as $\Tt(x_{\text{min}}=1) \sim  [\bar{\mu}/4- J]/q$, which is zero at the endpoint of the coexistence line $\bar{\mu} = 4 J$.

Together with the expression \eqref{xcutoff}, we have near the cutoff, so small $\delta = q^2(r-r_{\text{max}})$,
\begin{equation}
    f'(r) = q \pi \mu (1-\delta/2)^{-1/2} - q\pi \mu_0 \label{nearCutfp}
\end{equation}
which yields the scalar curvature $-f^{(2)}(r_{\text{max}})/2 = q^{3} \mu/4$.

Now using \eqref{Cp}, we find the scalar curvature $\Rr_2(r_{\text{H}}) =	-4\pi T/ C_{\mu}$. At the IR boundary $\Qq = 1/q$, $\beta\epsilon \to \pi v \tan(\pi v/2)$, for which \eqref{genC} reduces to
\begin{equation}
	q^2 C_\mu \sim  \frac{\pi^2 v}{b(v)} + \frac{4}{1+\beta\epsilon/q} \sim \frac{2\pi^2}{\beta J} + \frac{4}{1+\beta J/q}, \label{specHeat}
\end{equation}
where we have replaced $\beta\epsilon/q = \beta J/q + \Oo(T/q)$ and $v/b(v) \to 2 T/J$, We may evaluate this at the boundary using \eqref{specHeat} 
\begin{equation}
\Rr_2(r_{\text{max}}) \sim -4\pi q^{2}/\left[\frac{2\pi^2}{J} + \frac{4}{\Tt(r_{\text{max}})+J/q}\right],
\end{equation}
yielding 
\begin{equation}
\Rr_2(r_{\text{max}}) \sim -\pi q^{3/2}\tilde{\mu}/4, \label{boundaryCurv}
\end{equation}
Note that this is in agreement with the above result \eqref{nearCutfp}.

\section{Altered  v.s. standard SYK model \label{DavComp}}

    Here, we discuss the thermodynamic analysis of the \emph{altered } large-$q$ SYK model $H_{\text{alt}}(\beta\mu)$ introduced by Davison et al. in \cite[App. C: Large $q$ expansion of the SYK model]{Davison2017}.
    
    \begin{table}[ht]
    \centering
    \caption{Comparison between the standard \cite{Louw2023} and altered \cite[App. C]{Davison2017} SYK models.\hspace*{\fill} \label{tab:title2} } 
    \begin{tabular}{|c|c| c |} 
    \hline
    Model& Altered  SYK& Standard SYK\\
    \hline
    Hamiltonian & $H_{\text{alt}}(\beta\mu)$ \cite{Davison2017} & $\Hh$ \cite{Louw2023}
    \\
    Phase transition & No & Yes\\
    Interactions at $\Qq = \Oo(q^0)$ & Non-trivial & Trivial\\
    Ground state energy density & $\Oo(q^{-2})$ & $e^{-q\Qq^2} \Oo(q^{-2})$\\
    \hline
    \end{tabular}
    \end{table}
    
    Since the models are identical up to a choice in coupling, the self-energies are also equivalent. The difference arises in the effective coupling from \cite[C.8]{Davison2017} $\Jj(\Qq) = J(\beta\mu)/c(\beta\mu)$, where $c(\beta\mu) \equiv [2 \cosh(\beta\mu/2)]^{q/2-1} \sim  e^{2 q [\beta\mu/4]^2}$ for large $q$, for any $\beta\mu = \Oo(q^0)$. To counteract this one can rescale the bare coupling, which directly enters the Hamiltonian, as $J(\beta\mu) \to J_0 c(\beta\mu)$. This yields an altered Hamiltonian $H_{\text{alt}}(\beta\mu)$. This $\beta\mu$ dependent Hamiltonian drastically changes the thermodynamics of the standard temperature-independent SYK Hamiltonian with some differences listed in table \ref{tab:title2}. For instance, the lack of a negative compressibility \cite[C22]{Davison2017} in $H_{\text{alt}}(\beta\mu)$, is an indication that $H(\beta\mu)$ \emph{does not} have a phase transition. In contrast, the unaltered Hamiltonian \eqref{H} has a quantitatively and qualitatively similar phase diagram to its finite $q$ equivalents \cite{Azeyanagi2018Feb,Ferrari2019Jul,Louw2023}.

\end{document}